\documentclass[twocolumn,trackchanges]{aastex701}

\begin{document}

\title{Systematic Study of the Simultaneous Events Detected by GECAM}

\correspondingauthor{Feng-Rong Zhu, Shao-Lin Xiong}
\email{zhufr@home.swjtu.edu.cn, xiongsl@ihep.ac.cn}

\author{Yang-Zhao Ren}
\affil{School of Physical Science and Technology, Southwest Jiaotong University, Chengdu Sichuan, 611756, China}
\affil{State Key Laboratory of Particle Astrophysics, Institute of High Energy Physics, Chinese Academy of Sciences, 19B Yuquan Road, Beijing 100049, China}
\email{renyz@ihep.ac.cn}  

\author{Feng-Rong Zhu*} 
\affil{School of Physical Science and Technology, Southwest Jiaotong University, Chengdu Sichuan, 611756, China}
\affil{School of Science, Xizang University, Lhasa 850001, China}
\email[]{zhufr@home.swjtu.edu.cn}

\author[0000-0002-4771-7653]{Shao-Lin Xiong*}
\affil{State Key Laboratory of Particle Astrophysics, Institute of High Energy Physics, Chinese Academy of Sciences, 19B Yuquan Road, Beijing 100049, China}
\email[]{xiongsl@ihep.ac.cn}

\author{Yan-Qiu Zhang}
\affil{State Key Laboratory of Particle Astrophysics, Institute of High Energy Physics, Chinese Academy of Sciences, 19B Yuquan Road, Beijing 100049, China}
\affil{University of Chinese Academy of Sciences, Chinese Academy of Sciences, Beijing 100049, China}
\email{zhangyanqiu@ihep.ac.cn}

\author{Chen-Wei Wang}
\affil{State Key Laboratory of Particle Astrophysics, Institute of High Energy Physics, Chinese Academy of Sciences, 19B Yuquan Road, Beijing 100049, China}
\affil{University of Chinese Academy of Sciences, Chinese Academy of Sciences, Beijing 100049, China}
\email{cwwang@ihep.ac.cn}

\author{Jia-cong Liu}
\affil{State Key Laboratory of Particle Astrophysics, Institute of High Energy Physics, Chinese Academy of Sciences, 19B Yuquan Road, Beijing 100049, China}
\affil{University of Chinese Academy of Sciences, Chinese Academy of Sciences, Beijing 100049, China}
\email{liujc98@ihep.ac.cn}

\author{Hao-xuan Guo}
\affil{State Key Laboratory of Particle Astrophysics, Institute of High Energy Physics, Chinese Academy of Sciences, 19B Yuquan Road, Beijing 100049, China}
\affil{Department of Nuclear Science and Technology, School of Energy and Power Engineering, Xi’an Jiaotong University, Xi’an, China}
\email{guohx@ihep.ac.cn}

\author{Shuo Xiao}
\affil{School of Physics and Electronic Science, Guizhou Normal University, Guiyang 550001, China}
\affil{Guizhou Provincial Key Laboratory of Radio Astronomy and Data Processing, Guizhou Normal University, \\Guiyang 550001, China}
\email{xiaoshuo@gznu.edu.cn}

\author[]{Dong-Ya Guo}
\affil{State Key Laboratory of Particle Astrophysics, Institute of High Energy Physics, Chinese Academy of Sciences, 19B Yuquan Road, Beijing 100049, China}
\email{guody@ihep.ac.cn}

\author[]{Zheng-Hua An}
\affil{State Key Laboratory of Particle Astrophysics, Institute of High Energy Physics, Chinese Academy of Sciences, 19B Yuquan Road, Beijing 100049, China}
\email{anzh@ihep.ac.cn}

\author{Ce Cai}
\affil{College of Physics and Hebei Key Laboratory of Photophysics Research and Application, 
\\Hebei Normal University, Shijiazhuang, Hebei 050024, China}
\affil{Shijiazhuang Key Laboratory of Astronomy and Space Science, Hebei Normal University, Shijiazhuang, Hebei 050024, China}
\email{caice@hebtu.edu.cn}

\author[0000-0002-8097-3616]{Pei-Yi Feng}
\affil{State Key Laboratory of Particle Astrophysics, Institute of High Energy Physics, Chinese Academy of Sciences, 19B Yuquan Road, Beijing 100049, China}
\email{fengpeiyi@ihep.ac.cn}

\author[]{Min Gao}
\affil{State Key Laboratory of Particle Astrophysics, Institute of High Energy Physics, Chinese Academy of Sciences, 19B Yuquan Road, Beijing 100049, China}
\email{gaom@ihep.ac.cn}

\author[]{Ke Gong}
\affil{State Key Laboratory of Particle Astrophysics, Institute of High Energy Physics, Chinese Academy of Sciences, 19B Yuquan Road, Beijing 100049, China}
\email{gongk@ihep.ac.cn}

\author[]{Yue Huang}
\affil{State Key Laboratory of Particle Astrophysics, Institute of High Energy Physics, Chinese Academy of Sciences, 19B Yuquan Road, Beijing 100049, China}
\email{huangyue@ihep.ac.cn}

\author[]{Bing Li}
\affil{State Key Laboratory of Particle Astrophysics, Institute of High Energy Physics, Chinese Academy of Sciences, 19B Yuquan Road, Beijing 100049, China}
\email{libing@ihep.ac.cn}

\author{Xiao-Bo Li}
\affil{State Key Laboratory of Particle Astrophysics, Institute of High Energy Physics, Chinese Academy of Sciences, 19B Yuquan Road, Beijing 100049, China}
\email{lixb@ihep.ac.cn}

\author[]{Xin-Qiao Li}
\affil{State Key Laboratory of Particle Astrophysics, Institute of High Energy Physics, Chinese Academy of Sciences, 19B Yuquan Road, Beijing 100049, China}
\email{lixq@ihep.ac.cn}

\author[]{Xiao-Jing Liu}
\affil{State Key Laboratory of Particle Astrophysics, Institute of High Energy Physics, Chinese Academy of Sciences, 19B Yuquan Road, Beijing 100049, China}
\email{liuxj@ihep.ac.cn}

\author[]{Ya-Qing Liu}
\affil{State Key Laboratory of Particle Astrophysics, Institute of High Energy Physics, Chinese Academy of Sciences, 19B Yuquan Road, Beijing 100049, China}
\email{liuyaqing@ihep.ac.cn}

\author[]{Xiang Ma}
\affil{State Key Laboratory of Particle Astrophysics, Institute of High Energy Physics, Chinese Academy of Sciences, 19B Yuquan Road, Beijing 100049, China}
\email{max@ihep.ac.cn}

\author[]{Wen-Xi Peng}
\affil{State Key Laboratory of Particle Astrophysics, Institute of High Energy Physics, Chinese Academy of Sciences, 19B Yuquan Road, Beijing 100049, China}
\email{pengwx@ihep.ac.cn}

\author[]{Rui Qiao}
\affil{State Key Laboratory of Particle Astrophysics, Institute of High Energy Physics, Chinese Academy of Sciences, 19B Yuquan Road, Beijing 100049, China}
\email{qiaorui@ihep.ac.cn}

\author[]{Li-Ming Song}
\affil{State Key Laboratory of Particle Astrophysics, Institute of High Energy Physics, Chinese Academy of Sciences, 19B Yuquan Road, Beijing 100049, China}
\email{songlm@ihep.ac.cn}

\author[]{Xi-Lei Sun}
\affil{State Key Laboratory of Particle Astrophysics, Institute of High Energy Physics, Chinese Academy of Sciences, 19B Yuquan Road, Beijing 100049, China}
\email{sunxl@ihep.ac.cn}

\author{Wen-Jun Tan}
\affil{State Key Laboratory of Particle Astrophysics, Institute of High Energy Physics, Chinese Academy of Sciences, 19B Yuquan Road, Beijing 100049, China}
\affil{University of Chinese Academy of Sciences, Chinese Academy of Sciences, Beijing 100049, China}
\email{tanwj@ihep.ac.cn}

\author[]{Jin Wang}
\affil{State Key Laboratory of Particle Astrophysics, Institute of High Energy Physics, Chinese Academy of Sciences, 19B Yuquan Road, Beijing 100049, China}
\email{jinwang@ihep.ac.cn}

\author[]{Jin-Zhou Wang}
\affil{State Key Laboratory of Particle Astrophysics, Institute of High Energy Physics, Chinese Academy of Sciences, 19B Yuquan Road, Beijing 100049, China}
\email{jzwang@ihep.ac.cn}

\author{Ping Wang}
\affil{State Key Laboratory of Particle Astrophysics, Institute of High Energy Physics, Chinese Academy of Sciences, 19B Yuquan Road, Beijing 100049, China}
\email{pwang@ihep.ac.cn}

\author{Yue Wang}
\affil{State Key Laboratory of Particle Astrophysics, Institute of High Energy Physics, Chinese Academy of Sciences, 19B Yuquan Road, Beijing 100049, China}
\affil{University of Chinese Academy of Sciences, Chinese Academy of Sciences, Beijing 100049, China}
\email{yuewang@ihep.ac.cn}

\author[]{Xiang-Yang Wen}
\affil{State Key Laboratory of Particle Astrophysics, Institute of High Energy Physics, Chinese Academy of Sciences, 19B Yuquan Road, Beijing 100049, China}
\email{wenxy@ihep.ac.cn}

\author{Sheng-Lun Xie}
\affil{State Key Laboratory of Particle Astrophysics, Institute of High Energy Physics, Chinese Academy of Sciences, 19B Yuquan Road, Beijing 100049, China}
\affil{Institute of Astrophysics, Central China Normal University, Wuhan 430079, China}
\email{xiesl@ihep.ac.cn}

\author{Wang-Chen Xue}
\affil{State Key Laboratory of Particle Astrophysics, Institute of High Energy Physics, Chinese Academy of Sciences, 19B Yuquan Road, Beijing 100049, China}
\affil{University of Chinese Academy of Sciences, Chinese Academy of Sciences, Beijing 100049, China}
\email{xuewc@ihep.ac.cn}

\author[]{Sheng Yang}
\affil{State Key Laboratory of Particle Astrophysics, Institute of High Energy Physics, Chinese Academy of Sciences, 19B Yuquan Road, Beijing 100049, China}
\email{yangsheng@ihep.ac.cn}

\author[]{Qi-Bin Yi}
\affil{State Key Laboratory of Particle Astrophysics, Institute of High Energy Physics, Chinese Academy of Sciences, 19B Yuquan Road, Beijing 100049, China}
\affil{School of Physics and Optoelectronics, Xiangtan University, Xiangtan 411105, China}
\affil{Key Laboratory of Lithium Battery New Energy Materials and Devices of Jiangxi Education Department, College of Intelligent Manufacturing and Materials \& Chemical Engineering, Yichun University, Yichun, Jiangxi Province, 336000, China}
 \email{yiqb@ihep.ac.cn}

\author{Zheng-Hang Yu}
\affil{State Key Laboratory of Particle Astrophysics, Institute of High Energy Physics, Chinese Academy of Sciences, 19B Yuquan Road, Beijing 100049, China}
\affil{University of Chinese Academy of Sciences, Chinese Academy of Sciences, Beijing 100049, China}
\email{zhyu@ihep.ac.cn}

\author[]{Da-Li Zhang}
\affil{State Key Laboratory of Particle Astrophysics, Institute of High Energy Physics, Chinese Academy of Sciences, 19B Yuquan Road, Beijing 100049, China}
\email{zhangdl@ihep.ac.cn}

\author[]{Fan Zhang}
\affil{State Key Laboratory of Particle Astrophysics, Institute of High Energy Physics, Chinese Academy of Sciences, 19B Yuquan Road, Beijing 100049, China}
\email{zhangfan@ihep.ac.cn}

\author[]{Hong-Mei Zhang}
\affil{Computing Center, Institute of High Energy Physics, Chinese Academy of Sciences, 19B Yuquan Road, Beijing 100049, China}
\email{zhanghm@ihep.ac.cn}

\author{Jin-Peng Zhang}
\affil{State Key Laboratory of Particle Astrophysics, Institute of High Energy Physics, Chinese Academy of Sciences, 19B Yuquan Road, Beijing 100049, China}
\affil{University of Chinese Academy of Sciences, Chinese Academy of Sciences, Beijing 100049, China}
\email{zhangjinpeng@ihep.ac.cn}

\author{Peng Zhang}
\affil{State Key Laboratory of Particle Astrophysics, Institute of High Energy Physics, Chinese Academy of Sciences, 19B Yuquan Road, Beijing 100049, China}
\affil{College of Electronic and Information Engineering, Tongji University, Shanghai 201804, China}
\email{zhangp97@ihep.ac.cn}

\author{Shuang-Nan Zhang}
\affil{State Key Laboratory of Particle Astrophysics, Institute of High Energy Physics, Chinese Academy of Sciences, 19B Yuquan Road, Beijing 100049, China}
\email{zhangsn@ihep.ac.cn}

\author{Wen-Long Zhang}
\affil{State Key Laboratory of Particle Astrophysics, Institute of High Energy Physics, Chinese Academy of Sciences, 19B Yuquan Road, Beijing 100049, China}
\affil{School of Physics and Physical Engineering, Qufu Normal University, Qufu, Shandong 273165, China}
\email{zhangwl@ihep.ac.cn}

\author{Zhen Zhang}
\affil{State Key Laboratory of Particle Astrophysics, Institute of High Energy Physics, Chinese Academy of Sciences, 19B Yuquan Road, Beijing 100049, China}
\email{zhangzhen@ihep.ac.cn}

\author[]{Xiao-Yun Zhao}
\affil{State Key Laboratory of Particle Astrophysics, Institute of High Energy Physics, Chinese Academy of Sciences, 19B Yuquan Road, Beijing 100049, China}
\email{xyzhao@ihep.ac.cn}

\author{Yi Zhao}
\affil{School of Computer and Information, Dezhou University, Dezhou Shandong, 253023, China}
\email{yizhao@dzu.edu.cn}

\author{Chao Zheng}
\affil{State Key Laboratory of Particle Astrophysics, Institute of High Energy Physics, Chinese Academy of Sciences, 19B Yuquan Road, Beijing 100049, China}
\affil{University of Chinese Academy of Sciences, Chinese Academy of Sciences, Beijing 100049, China}
\email{zhengchao97@ihep.ac.cn}

\author{Shi-Jie Zheng}
\affil{State Key Laboratory of Particle Astrophysics, Institute of High Energy Physics, Chinese Academy of Sciences, 19B Yuquan Road, Beijing 100049, China}
\email{zhengsj@ihep.ac.cn}



\begin{abstract}
GECAM is a constellation of all-sky monitors in hard X-ray and gamma-ray band primarily aimed at high energy transients such as gamma-ray bursts, soft gamma-ray repeaters, solar flares and terrestrial gamma-ray flashes. As GECAM has the highest temporal resolution (0.1~$\mu$s) among instruments of its kind, it can identify the so-called simultaneous events (STE) that deposit signals in multiple detectors nearly at the same time (with a 0.3~$\mu$s window). However, the properties and origin of STE have not yet been explored. 
In this work, we implemented, for the first time, a comprehensive analysis of the STE detected by GECAM, including their morphology, energy deposition, and the dependence on the geomagnetic coordinates. We find that these STE probably result from direct interactions between high-energy charged cosmic rays and satellite. These results demonstrate that GECAM can detect, identify, and characterize high-energy cosmic rays, making it a Micro Cosmic-Ray Observatory (MICRO) in low Earth orbit. 
\end{abstract}



\section{Introduction}\label{section1} 

The first joint observation of a gamma-ray burst (GRB 170817A) and a gravitational wave (GW170817) in 2017 heralded the dawn of multi-messenger gravitational-wave astronomy \citep{Abbott2017, Goldstein2017, LiT2018}. Motivated by and designed for the detection of gamma-ray burst associated with gravitational waves \citep{XiongS2020}, the Gravitational-Wave High-Energy Electromagnetic Counterpart All-Sky Monitor (GECAM) was proposed in 2016. 

Initially, GECAM comprises two microsatellites (GECAM-A and GECAM-B) in low-Earth orbit at an altitude of 600 km and an inclination of 29°, which were launched on December 10, 2020. Each GECAM satellite carries 25 gamma-ray detectors (GRDs) \citep{AnZ2022}, 8 charged-particle detectors (CPDs) \citep{XuY2022}, and an electronics box (EBOX) \citep{LiX2022}. The GRDs employ lanthanum bromide (LaBr$_3$) crystals coupled to silicon photomultipliers (SiPM) for gamma-ray detection, while the CPDs use plastic scintillators with SiPM readout to monitor charged particles and help GRDs to discriminate between gamma-ray bursts and particle events. The EBOX is located in the payload dome and 
is responsible for acquiring and processing detection data from the GRDs and CPDs, enabling on-board triggering and localization of gamma-ray transients. It also handles 
communication with the satellite platform and distribution of the secondary power supply. 

To date, GECAM has developed to a constellation consisting of GECAM-A/B, GECAM-C \citep{LiX2025} and GECAM-D \citep{FengP2024}. Thus, GECAM offers all-sky coverage, high sensitivity, in-flight localization, broad-energy range, and low-energy threshold; it can distribute real-time alerts of gamma-ray transients to guide follow-up observations. GECAM made many observation and obtained fruitful results \citep{WangC2024}, specifically, these results cover gamma-ray bursts \citep{Moradi2024,ZhangY2024}, magnetar bursts \citep{XieS2025}, solar flares \citep{SuY2020,ZhaoH2023}, terrestrial gamma-ray flashes, terrestrial electron beams and a new type of high energy transients in the orbit \citep{ZhaoY2023}.

A standout feature of GECAM is its unprecedented time resolution of $0.1\,\mu\mathrm{s}$ among 
wide-field gamma-ray monitors 
\citep{ZhangP2021,XiaoS2023,ZhaoY2023}. 
For comparison, Insight-HXMT’s High Energy (HE) instrument achieves a time resolution of $2\,\mu\mathrm{s}$ \citep{ZhangS2020,LiuC2020}, while AstroSat’s LAXPC instrument $10\,\mu\mathrm{s}$ \citep{Agrawal2006,Yadav2017}, Swift/BAT $100\,\mu\mathrm{s}$ \citep{Sakamoto2008}, Konus-Wind $2\,\mathrm{ms}$ \citep{Aptekar1995}, and INTEGRAL/SPI-ACS $50\,\mathrm{ms}$ \citep{vonKienlin2003}. High time resolution could facilitate precise source localization via time-delay methods \citep{XiaoS2021}, investigation of emission mechanisms through timing analysis \citep{Bernardini2015}, detection of high-frequency quasi-periodic oscillations in magnetars \citep{Huppenkothen2014,Roberts2023}, and spacecraft-based pulsar navigation \citep{ZhengS2019,XiaoS2023,LuoX2023}.
Importantly, high time resolution allows GECAM to identify the phenomena called simultaneous events (STE), which are signals\footnote{They are usually called events or triggers, which are output by the detector.} recorded by multiple detectors of GECAM within a very small time window (i.e. $0.3\,\mu\mathrm{s}$). 

Although instruments like Fermi/GBM previously studied simultaneous multi-detector events for cosmic-ray background for TGF search \citep{Briggs2013}—those studies were limited to $2\,\mu\mathrm{s}$ time resolution and lacked a systematic analysis.
Indeed, GECAM is the first gamma-ray monitor to compile a large, high time precision sample of STE. However, the properties and origin of STE has not systematically explored yet.

This work is motivated by the need to characterize the properties and probe the origin of these STE. Indeed, 
even if the cosmic-ray shower is the probable source of STE, there are (at least) two hypotheses which could account for the production of the cosmic-ray shower: (1) Satellite scenario, in which primary cosmic rays interact directly with satellite structures or even detector materials to induce particle cascades that simultaneously leave signals in multiple detectors. (2) Atmospheric scenario, in which primary cosmic rays interact with nitrogen and oxygen nuclei in the Earth’s atmosphere to produce secondary particles that reach satellite altitude and are simultaneously registered by multiple detectors.
The primary task of this work is to determine whether cosmic rays are the primary source of the STE and in which scenario mentioned above.

This paper is structured as following: Section~\ref{section2} describes the GECAM instruments, their operating modes, the data, and presents the basic characteristics and preliminary statistics of STE. Section~\ref{section3} details the random-coincidence test method for assessing the probability of multi-detector sub-microsecond events (i.e. STE). Section~\ref{section4} analyzes observations in the zenith pointing mode (fully anti-Earth–pointing), including geomagnetic latitude effects, geomagetic field-line orientation dependence, and spatial clustering. Section~\ref{section5} discusses observations in non-zenith pointing mode. Section~\ref{section6} compares the spectral evolution of single-detector and multi-detector STE to reveal their energy distribution. Section~\ref{section7} synthesizes these analyses to discuss the origin and mechanisms of STE. Section~\ref{section8} summarizes this study.


\section{Instrument and Data} \label{section2}
\subsection{GECAM}

\begin{figure}[ht!]
\centering
\includegraphics[width=0.99\linewidth]{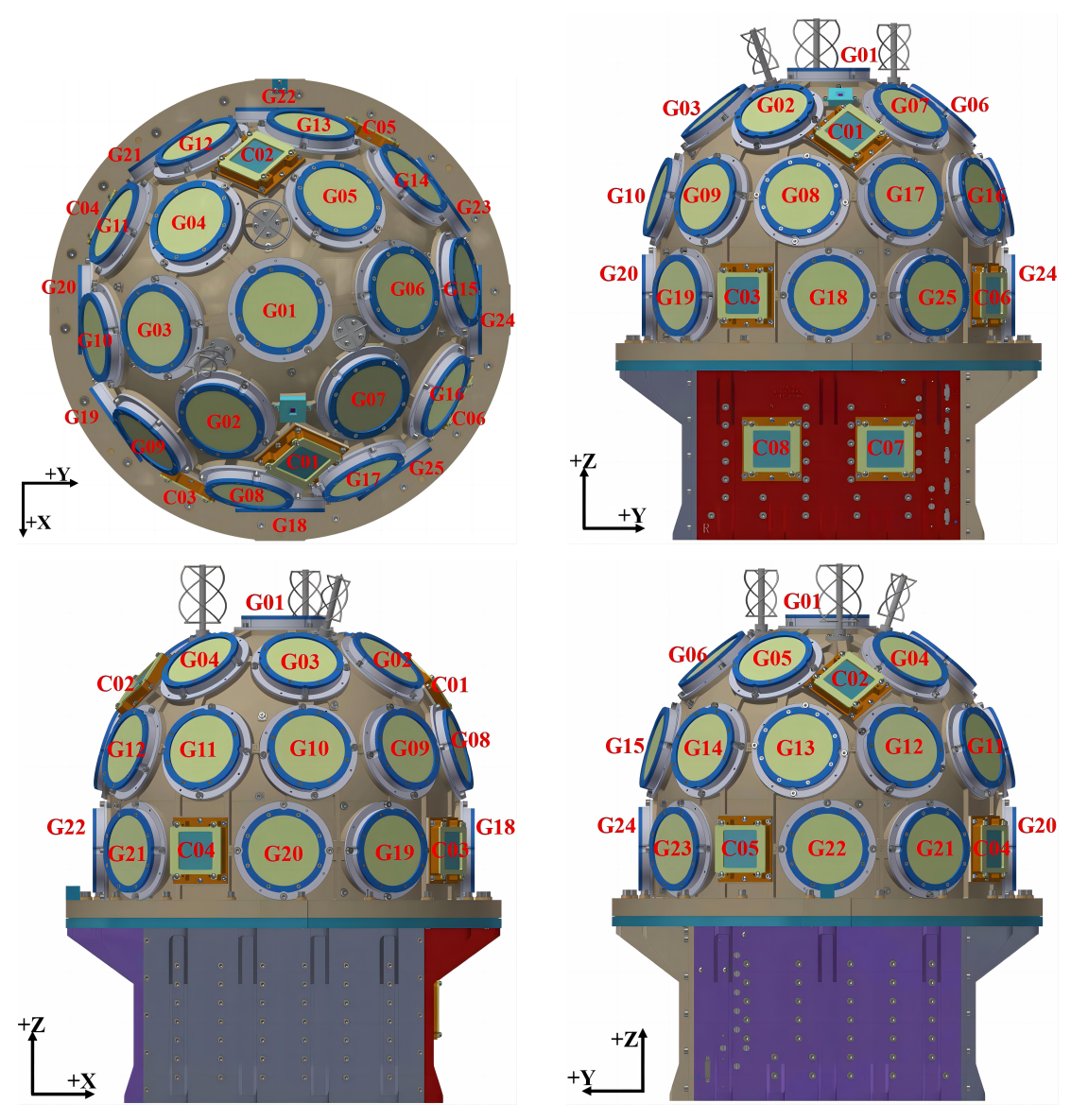}
\caption{The structural layout of the GECAM payload is shown from different perspectives and comprises 25 GRDs (designated G01 through G25) and eight CPDs (designated C01 through C08). The payload comprises two main components: the detector dome housing and the EBOX.The payload coordinate system is defined such that the $+X$, $+Y$, and $+Z$ axes align with the central normals of detectors G18, G24, and G01, respectively \citep{ZhaoYi2023}.}
\label{fig:fig1}
\end{figure}

The GECAM payload utilizes a multi‐detector array configuration, 25 GRDs and 6 CPDs are mounted on the satellite’s domed compartment and oriented in different directions for wide‐field sky monitoring, while the remaining 2 CPDs are affixed to the +X side of the EBOX. Each detector is numbered following the design specification to specify its spatial layout Figure \ref{fig:fig1}. Each GRD has a geometric area of approximately 45\,cm$^2$ (circular, 7.6\,cm diameter) and an on‐axis effective area of about 21\,cm$^2$ for 1\,MeV $\gamma$‐rays \citep{GuoD2020}. They detect high‐energy photons with energies ranging from approximately 15\,keV to 5\,MeV \citep{ZhangD2022}, employing a dual‐channel readout design: a high‐gain channel covering 15-300\,keV and a low‐gain channel covering 300\,keV–5\,MeV. The dead time for normal events is 4\,$\mu$s, while that for overflow events is not less than 69\,$\mu$s \citep{LiuY2022}. The CPDs have a geometric area of 16\,cm$^2$ (square, 4.0\,cm side length) and an on‐axis effective area of about 16\,cm$^2$ for 1\,MeV electrons \citep{XuY2022}. They are designed to measure charged‐particle flux variations in the 100\,keV–5\,MeV range \citep{ZhaoY2023}.

The GECAM detectors operate in two observational modes: normal mode and South Atlantic Anomaly (SAA) mode. 
In the normal mode, the detectors record event‐by‐event, with GRDs and CPDs recording the energy and arrival time of each incident event in real time. When the GRD count rate within several predefined time and energy windows significantly exceeds the background fluctuation, the payload automatically initiates an onboard real‐time analysis routine to determine the event’s trigger time, location, classification and intensity, and promptly downlinks this information to the ground station via the BeiDou satellite navigation system. Upon entering the SAA, the local charged‐particle flux sharply increases, causing a substantial rise in the data volume if keeping record of event‐by‐event data. Therefore the payload is set to automatically switch to SAA mode in which only a few detectors are kept on, the event-by-event data output is turned off, but the binned data recording is maintained. This SAA mode can not only control the data volume 
but also protect the detectors.

\begin{figure}[ht!]
\centering
\includegraphics[width=0.99\linewidth]{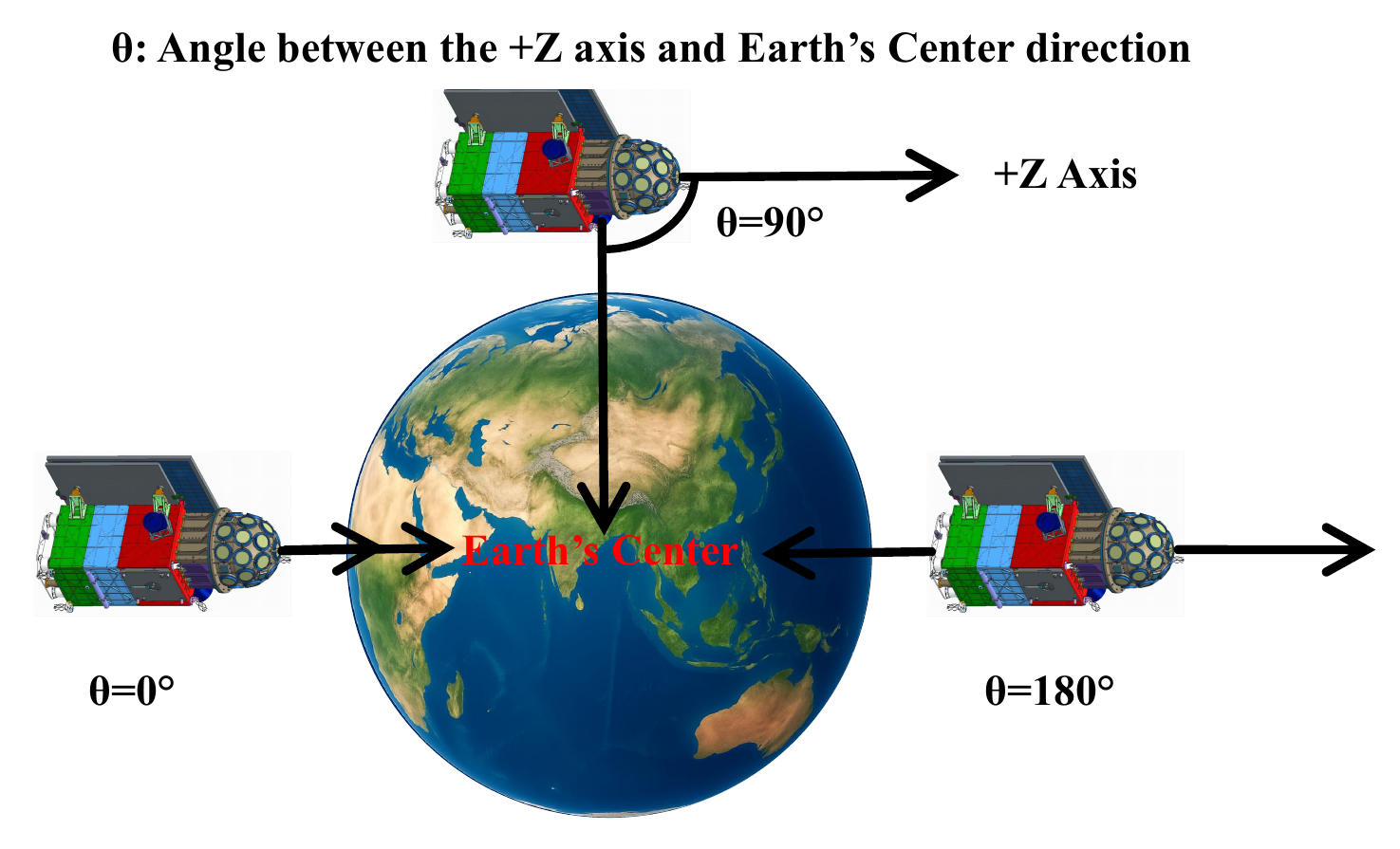}
\caption{Attitude variations of GECAM during its monitoring mission are shown. The angle \(\theta\) denotes the angle between the payload coordinate system’s \(+Z\) axis and the vector toward the Earth’s center. At \(\theta = 0^\circ\), GECAM observes toward the Earth. At \(\theta = 180^\circ\), GECAM observes directly away from the Earth. 
}
\label{fig:fig2}
\end{figure}

The GECAM satellite also works in two pointing modes: zenith observation or non-zenith observation. In the normal mode, the satellite conducts continuous all‐sky monitoring, with attitude adjustments made according to the observation plan. 
We define the angle $\theta$ between the payload boresight ($+Z$ axis) and the direction toward Earth’s center ($0^\circ \le \theta \le 180^\circ$) to quantify observation attitude. 
The payload attitude can be adjusted from zenith pointing to non‐zenith pointing, as shown in Figure \ref{fig:fig2}. The observation mode is determined by the preset observation plan. 

It is worth noting that to comprehensively investigate the source characteristics of STE, it is necessary to classify observational data according to satellite pointing modes. In the zenith-pointing mode, where the detector points vertically upward toward space, Earth occultation of detector's FOV is minimized, enabling a pure investigation of how geomagnetic latitude modulates STE. In the non-zenith pointing mode, by varying the satellite attitude from Earth-pointing to deep-space pointing orientations, we can test whether STE exhibit directional dependence, thereby determining whether they are sourced from specific sky directions. 

\subsection{Data}

GECAM data products follow a tiered classification in accordance with standard space astronomy conventions \citep{ZhengS2024}. Remarkably, dedicated scientific data products have been designed and produced for STE. STE data products are divided into hourly segments. Each file covers a 100\,s time window corresponding to the final 100\,s of the preceding hour, recording detailed count information from any two or more of the 33 detectors that got events recorded simultaneously, thereby providing foundational and convenient data for the temporal and spectral analysis for these STE.

Due to power constraints on GECAM-A observation, this study only utilizes the event-by-event data (EVT data) 
by GECAM-B \citep{CaiC2025}. 
The dataset spans multiple orbital periods and covers a wide range of geomagnetic latitudes. To avoid spatial non-uniformity effects in the CPD detectors on the GECAM dome, we analyze only trigger data from 25 GRDs. 

\begin{figure}[ht!]
\centering
\includegraphics[width=0.99\linewidth]{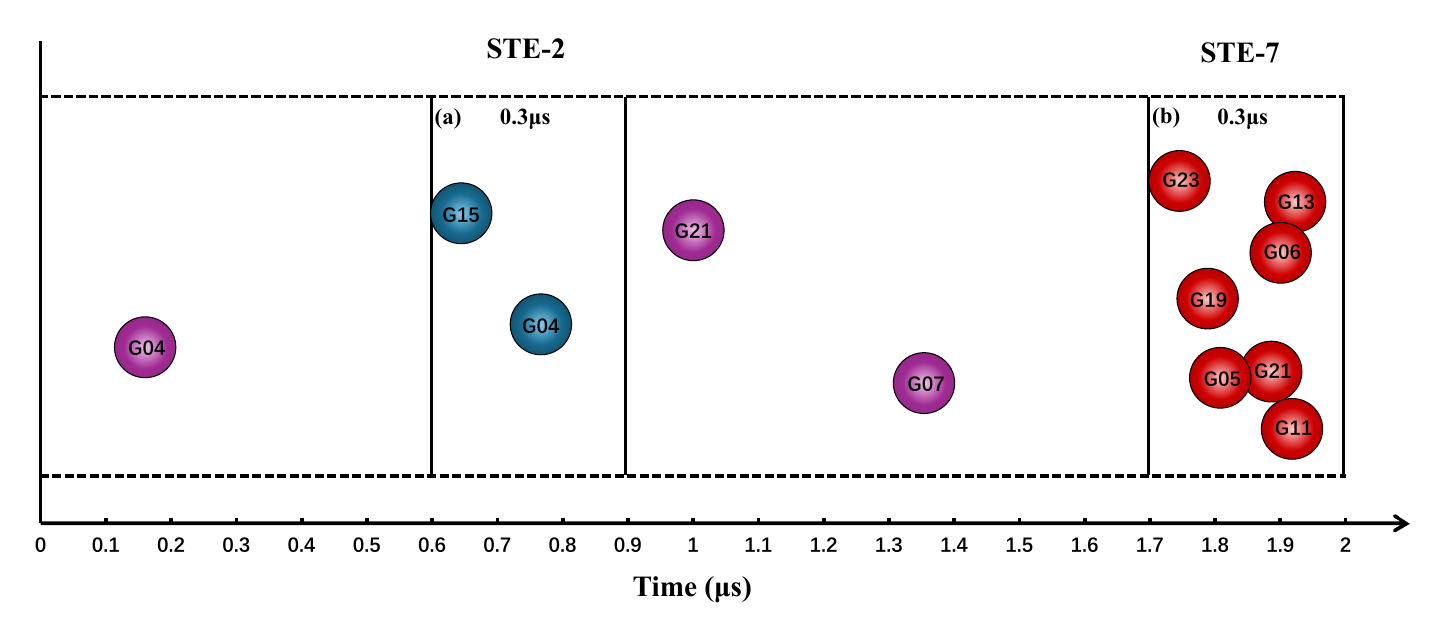}
\caption{Schematic diagram illustrating STE. Within a $2\,\mu\mathrm{s}$ interval, two-detector and seven-detector STE were detected in time windows (a) and (b), respectively. Outside these windows, no events satisfy the simultaneity criterion. Colored spheres indicate the triggered GRDs and their identifiers.}
\label{fig:fig3}
\end{figure}

The definition of STE is illustrated in Figure \ref{fig:fig3}. The time window \(\Delta t\) to identify STE is defined as \(0.3\,\mu\mathrm{s}\). A STE occurs when at least two detectors register events within this interval. 
This time window was determined through ground calibration with a $^{22}\mathrm{Na}$ source and on-orbit secondary cosmic-ray measurements to ensure that physically-related events (e.g. Compton scattering events, cosmic-ray shower events) could be accurately identified considering the fluctuation in the timing system among GECAM detectors \citep{XiaoS2022}. 

For each STE event, the number of triggered detectors, $N$, ranges from 2 to 25 (the total number of GRDs on board). For simplicity, STE with $N$ triggered detectors are called $N$-fold STE or simply dubbed as STE-$N$ throughout this paper. Analysis of the observed data reveals a large number of STE within a single hour, as shown in Figure \ref{fig:fig4}. 

\section{Coincidence effect on STE} \label{section3}

We note that, almost all STE should correspond to real physical events leaving multiple simultaneous signals in multiple detectors, such as the cosmic ray shower and Compton scattering, since the probability of multiple detectors being triggered within such a narrow window (0.3 $\mu$s) by pure random background of detectors is vanishingly small. Here we employ GECAM’s on-orbit background count rates and Poisson statistics to demonstrate this.

To verify the physical source, we let in the theoretical calculation the mean count rate of the $i$th detector in time bin $j$ be $\lambda_{i,j}$. Within a window of width $\Delta t = 0.3\,\mu\mathrm{s}$, the probability that a single detector triggers at least once is given by:
\begin{equation}
p_{i,j} \;=\; 1 - \exp\bigl(-\lambda_{i,j}\,\Delta t\bigr),
\end{equation}
Treating $X_{i,j} \sim \mathrm{Bernoulli}(p_{i,j})$ as independent random variables, the number of detectors triggered simultaneously, 
\begin{equation}
K = \sum_{i=1}^N X_{i,j},
\end{equation}
follows a Poisson–binomial distribution. Equivalently,
\begin{equation}
P_{j}(K = k) \;=\; 
\sum_{A \subset \{1,\dots,N\} \atop |A| = k}
\left(\prod_{i \in A} p_{i,j}\right)
\left(\prod_{i \notin A} (1 - p_{i,j})\right)\,.
\end{equation}
Recursive convolution is used to extract the polynomial coefficients efficiently. Averaging $P_{j}(K = k)$ over all time bins yields $\bar{P}_{\mathrm{th}}(k)$. In the Monte Carlo simulation, for each time bin $j$ and detector $i$, we sample 
\begin{equation}
n_{i,j} \sim \mathrm{Poisson}(\lambda_{i,j}\,\Delta t).
\end{equation}
If $n_{i,j} > 0$, the detector is considered triggered (i.e. record a signal). Repeating this procedure over all time bins and averaging gives $\bar{P}_{\mathrm{sim}}(k)$. Under the assumption of independent detector counts following Poisson processes, the Poisson–binomial model provides theoretical predictions for the $N$-fold random‐coincidence rate. Monte Carlo simulations furnish numerical confirmation of these predictions. The results show that the probability of $\ge 5$ detectors triggering within $0.3\,\mu\mathrm{s}$ due to random counts is extremely low (e.g., $2.24\times 10^{-14}$ for $N=5$). The theoretical predictions are displayed in Figure \ref{fig:fig5}, because Monte Carlo sampling is inherently inefficient for extremely low‐probability events, even after hundreds of millions of trials only the $k=2$ and $k=3$ orders yield nonzero, statistically significant random‐coincidence probabilities, whereas for $k\ge4$ virtually no events are sampled. Although such events are exceedingly rare, they are nonetheless observed in the GECAM data. 

\begin{figure}[ht!]
\centering
\includegraphics[width=0.99\linewidth]{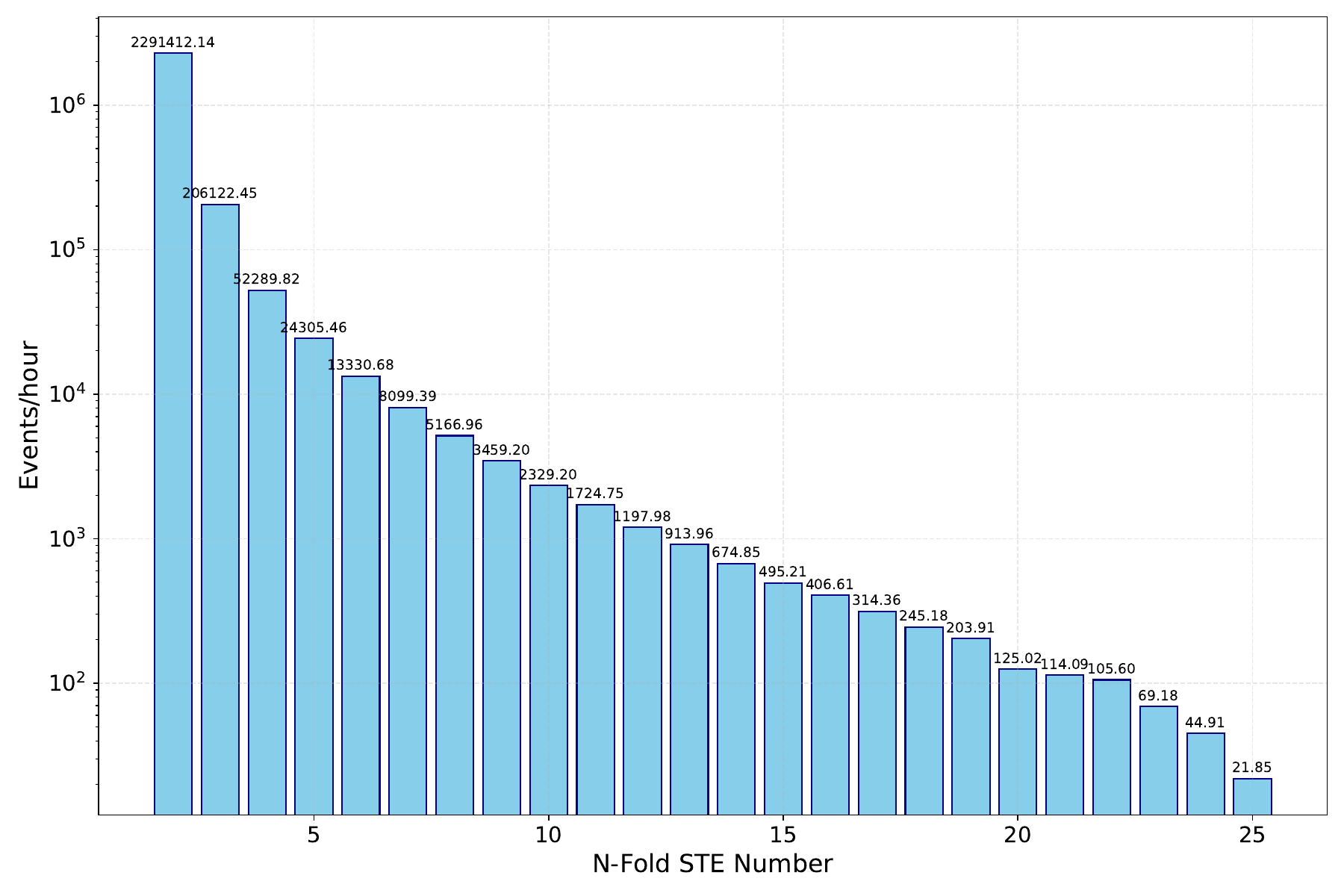}
\caption{Statistical distribution of STE observed on-orbit by GECAM-B. The figure illustrates the average number of STE per hour for varying numbers of detectors triggered ($N$), depicting the observational characteristics of events that simultaneously trigger multiple detectors.}
\label{fig:fig4}
\end{figure}

\begin{figure}[ht!]
\centering
\includegraphics[width=0.99\linewidth]{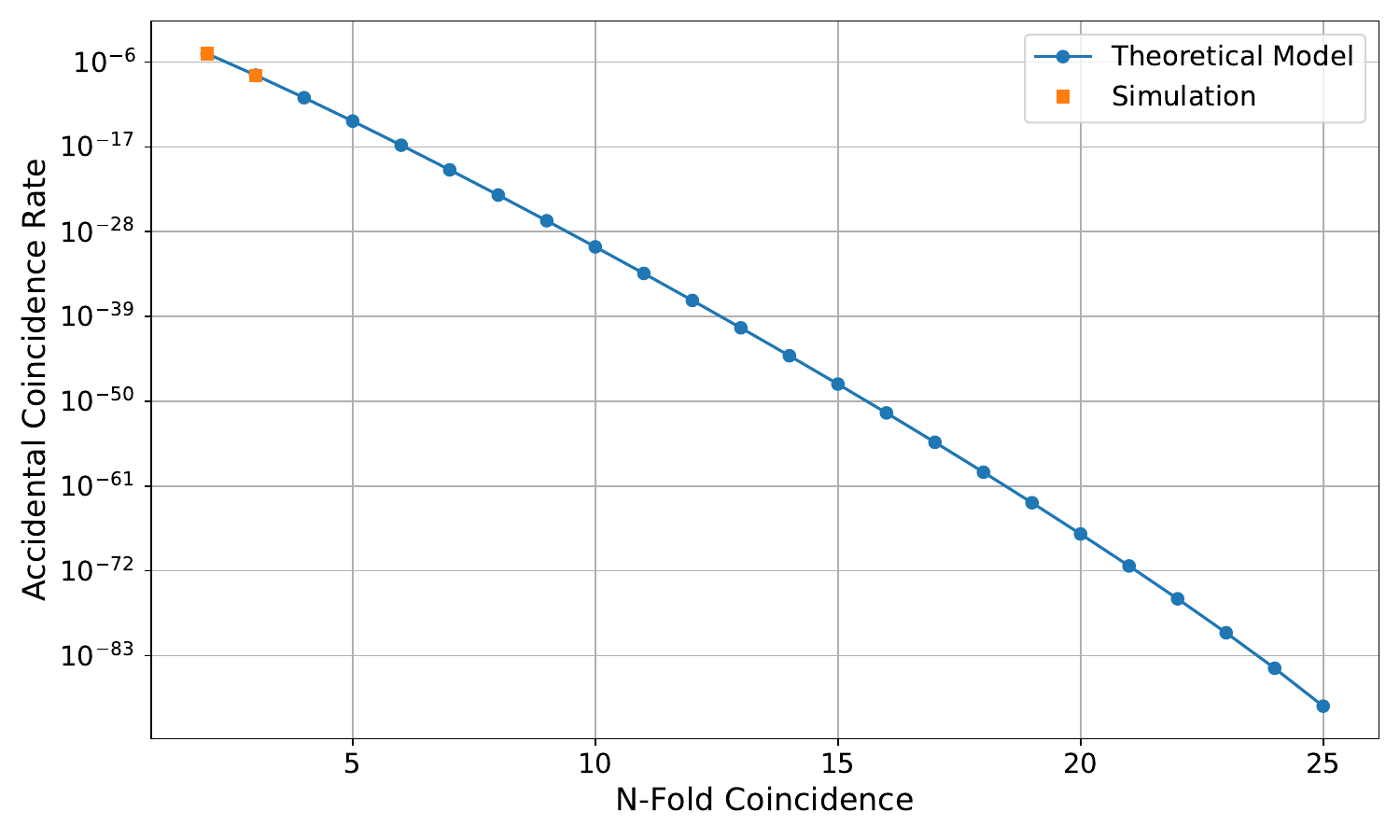}
\caption{Multi-detector random coincidence probability theoretical calculations (blue circles) and simulation results (orange squares) within a $0.3\,\mu\mathrm{s}$ time window.}
\label{fig:fig5}
\end{figure}

\begin{figure}[ht!]
\centering
\includegraphics[width=0.99\linewidth]{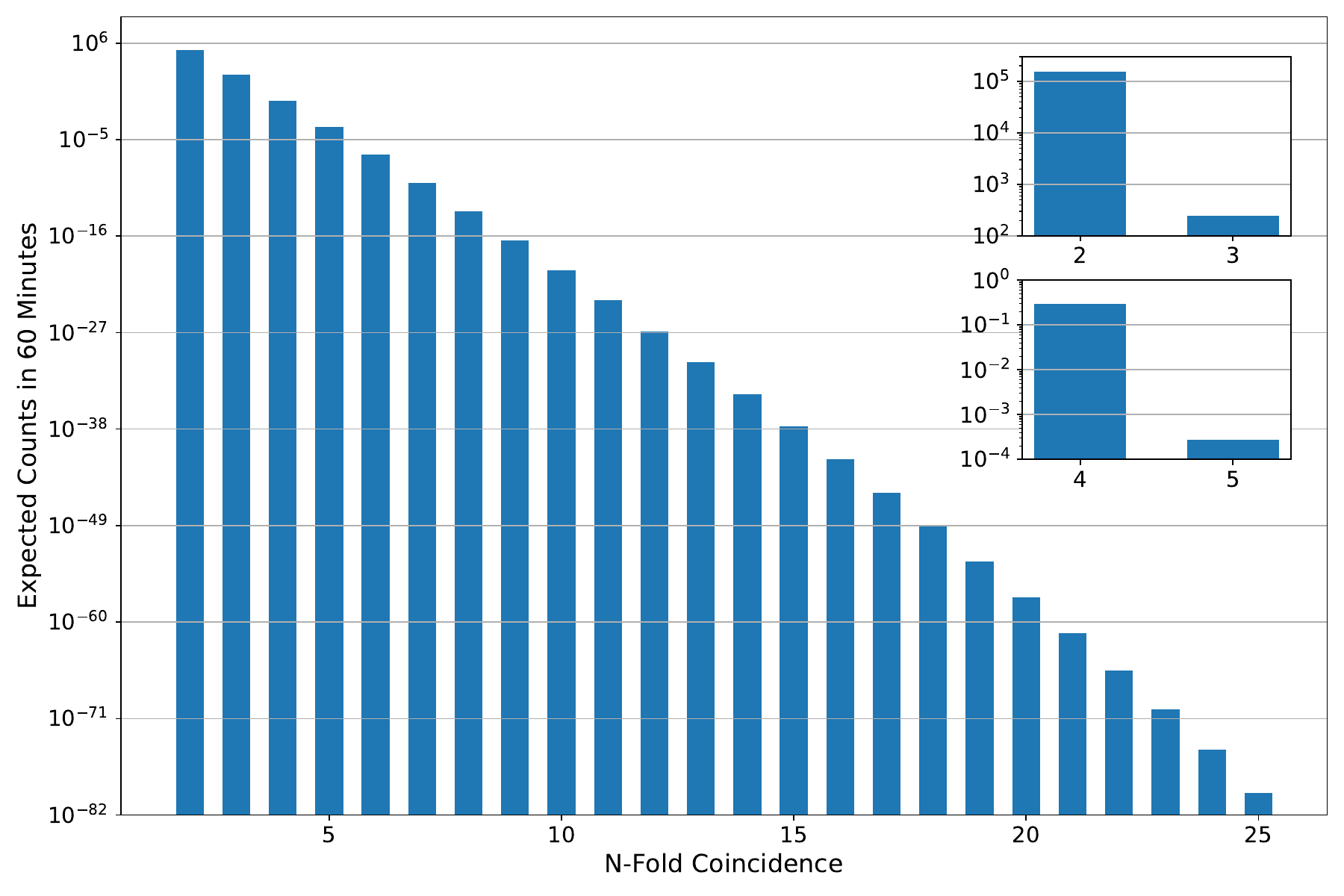}
\caption{Comparison plot of the expected random coincidence probability versus the number of triggered detectors (\(N\)) within one hour. The inset shows an enlarged view of the random coincidence probabilities for \(N=2,\,3,\,4,\,5\).}
\label{fig:fig6}
\end{figure}

Figure \ref{fig:fig6} plots the expected random‐coincidence probability within one hour as a function of $N$, comparison shows that the on-orbit measurements in Figure \ref{fig:fig4} lie well above the predicted random-coincidence background. For $N=5$, a random coincidence occurs only once every six months on average, implying that STE with $N\ge5$ cannot be attributed to chance and must share a common physical source. Notably, both GRDs and CPDs employ fully independent signal‐acquisition architectures \citep{LiuY2022}. Signals are routed through independent channels into the payload EBOX. The EBOX unifies data acquisition, processing, and transmission and provides bias‐voltage regulation for gain control, thus precluding physical crosstalk \citep{LiX2021}. 

To further investigate the composition of the GECAM-recorded STE, we take STE-5 as an illustrative example. Several possibilities merit discussion; for instance, four detectors may be genuinely triggered by a single physical particle while an unrelated accidental hit falls within the same $0.3 \,\mu\mathrm{s}$ window. Using Bayesian inference, we calculate the posterior probability of having $k$ accidental triggers among the observed five simultaneous triggers, based on the accidental trigger probabilities in each time bin and the prior distribution of the number of genuine triggers. The calculated conditional probabilities of accidental contamination among the observed STE-5 are $4.30\times10^{-3}$ for exactly one accidental hit, $9.82\times10^{-6}$ for two, $1.59\times10^{-8}$ for three, and $2.05\times10^{-11}$ for four, indicating that STE-5 are overwhelmingly pure with negligible accidental coincidence events from background.

\section{Zenith Pointing Observation Mode}\label{section4}
\subsection{Correlation with Geomagnetic Latitude}

\begin{figure*}[ht!]
\centering
\includegraphics[width=0.99\linewidth]{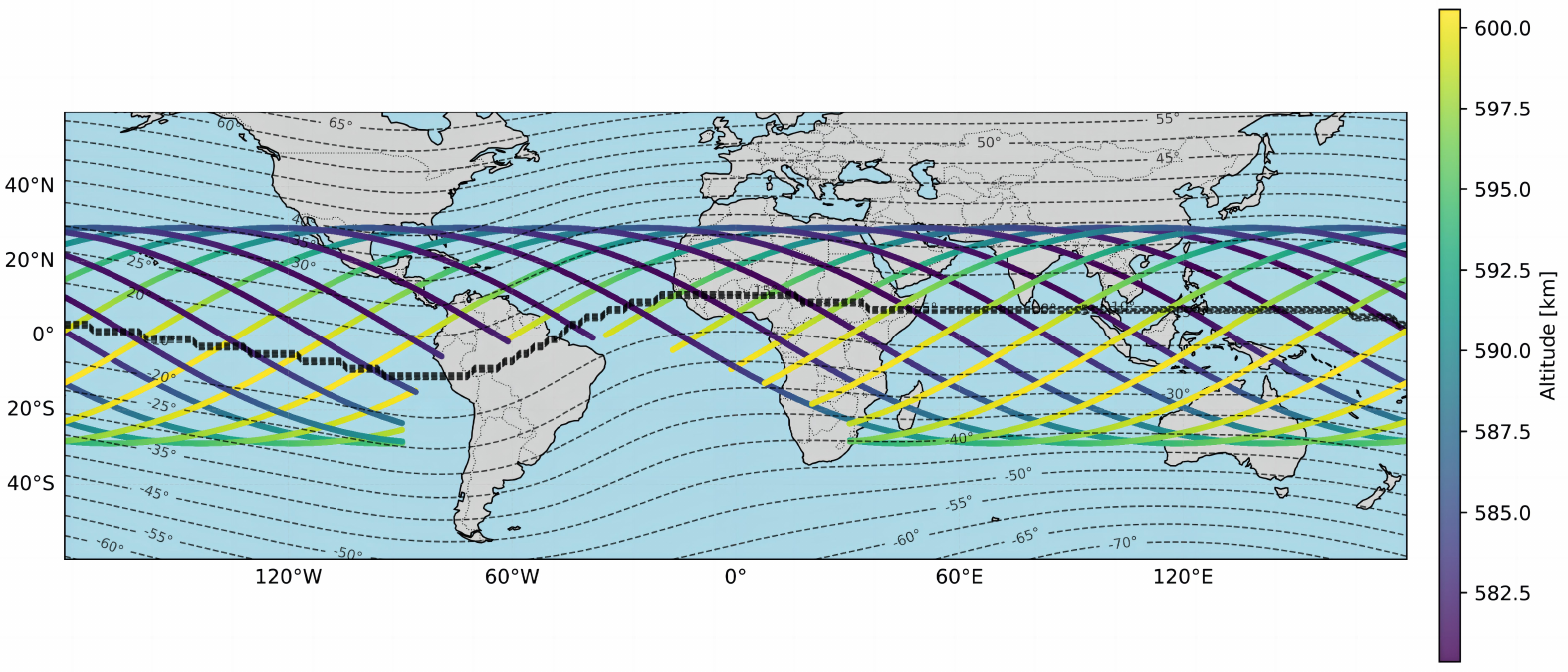}
\caption{The distribution map of the sub-satellite point trajectories of STE detected by GECAM-B during on-orbit operation, overlaid with geomagnetic latitude, where the lines represent sub-satellite point trajectories (excluding the SAA region), the colors indicate orbital altitude, and the black dashed lines denote geomagnetic latitude contours.}
\label{fig:fig7}
\end{figure*}

To investigate the spatial distribution of STE, we performed systematic analyses using both geographic and geomagnetic latitude as classification parameters. Statistics based on geographic latitude exhibit no obvious spatial pattern, indicating that the conventional geographic coordinate system does not effectively reveal features in STE distribution. When data are categorized by geomagnetic latitude, however, a clear correlation emerges. We converted geographic coordinates to geomagnetic coordinates using the standard Altitude-Adjusted Corrected Geomagnetic Coordinates (AACGM) model \citep{Shepherd2014} and examined the distribution of the sub-satellite point trajectories of STE detected by GECAM-B in geomagnetic latitude during on-orbit operations, details are displayed in Figure \ref{fig:fig7}. In the equatorial region, the thick black curve denotes the overlapping projection of the 0°–15° geomagnetic-latitude band at the satellite’s altitude. The geomagnetic axis is inclined by approximately 10° relative to the Earth’s rotation axis \citep{Macmillan2010}. As a result, the geomagnetic equator is displaced from the geographic equator and exhibits slight undulations with longitude. Consequently, points on the geographic equator map to roughly 0°–15° geomagnetic latitude during the transformation. Based on these geometric features, in the subsequent geomagnetic-latitude classification, on-orbit data within the ±40° geomagnetic-latitude range—but not including data within the ±15° band—were selected for statistical analysis.

\begin{figure*}[ht!]
\centering
\includegraphics[width=0.99\linewidth]{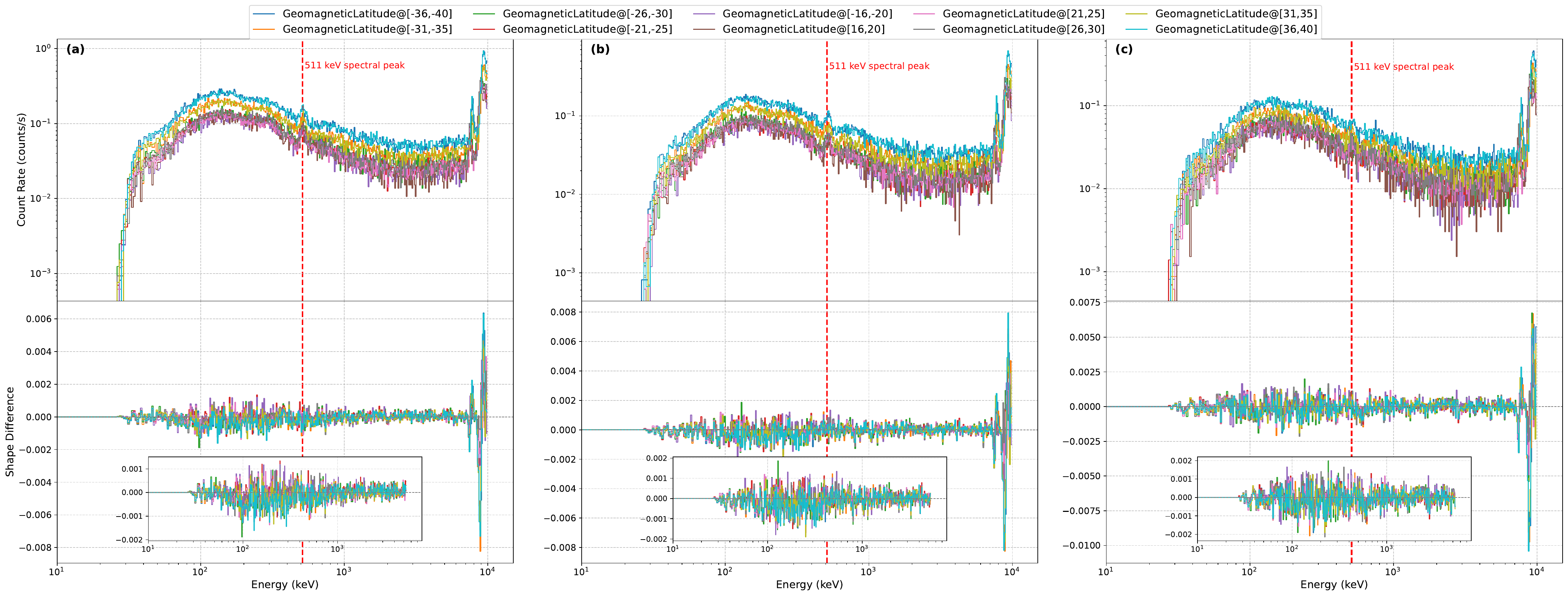}
\caption{Energy deposition spectra for the STE at different geomagnetic latitudes. Panels (a), (b), and (c) correspond to STE-5, STE-6, and STE-7, respectively. The upper panels show energy deposition spectra of STE at 10 different geomagnetic latitudes, while the lower panels compare the spectral shape differences. The red dashed line indicates the energy value at \(511\) keV.}
\label{fig:fig8}
\end{figure*}

\begin{figure}[ht!]
\centering
\includegraphics[width=0.99\linewidth]{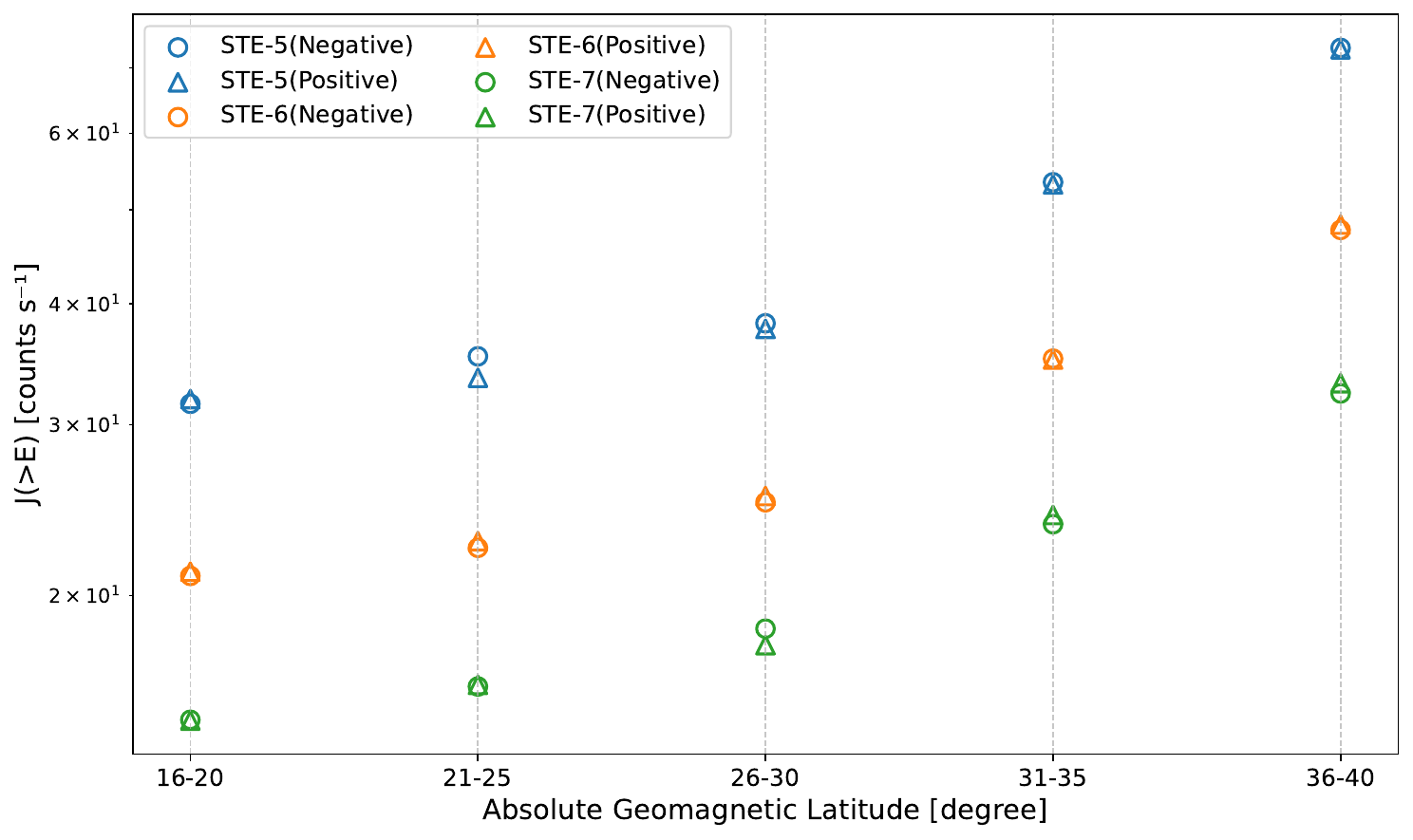}
\caption{Integrated flux distribution of STE across different geomagnetic latitude intervals. Blue, orange, and light green correspond, respectively, to the STE-5, STE-6, and STE-7 configurations.}
\label{fig:fig9}
\end{figure}

The geomagnetic modulation of STE becomes evident through their latitude-dependent energy spectra and integral fluxes. Energy-deposition spectra for STE-5, STE-6, and STE-7 were computed in 5°-wide geomagnetic-latitude bins, as displayed in Figure \ref{fig:fig8}. Spectra from magnetically symmetric intervals (e.g., [$-40^\circ$, $-36^\circ$] vs. [36°, 40°]) display highly consistent shapes, directly reflecting the north–south symmetry of Earth’s magnetic field and its modulation of cosmic‐ray transport. Using the spectrum in the [16°, 20°] bin as a reference, we normalize its probability‐density function to zero and apply the same normalization to all other bins to extract spectral “shape” parameters. The invariance of these shape parameters across geomagnetic latitudes indicates that the geomagnetic field primarily modulates the occurrence rate of STE, rather than altering their intrinsic energy‐distribution characteristics. The integral flux of STE-5, STE-6, and STE-7 increases monotonically with absolute geomagnetic latitude, as presented in Figure \ref{fig:fig9}. We note that this latitude-dependent enhancement is consistent with the geomagnetic‐rigidity cutoff effect. According to the magnetospheric transmission‐function theory \citep{Bobik2006}, cutoff rigidity decreases with increasing geomagnetic latitude, permitting more low‐energy cosmic rays to penetrate Earth’s magnetic shielding and reach satellite altitude, thereby boosting the observed primary cosmic‐ray flux at high latitudes. This modulation arises because only particles with momentum above the local cutoff rigidity can overcome the Lorentz force exerted by the geomagnetic field. We notice that, although the integral flux varies with latitude, the spectral shape remains stable, with differences confined to overall intensity and maximum energy, confirming that the geomagnetic field affects the event rate but not the fundamental physics of STE.

The energy spectra of STE exhibit a pronounced $511\,\mathrm{keV}$ positron‐annihilation line \citep{Siegert2016,GuoD2020}. This feature could arise from multiple processes: primary cosmic rays interacting with detector materials produce secondary particles and radioactive isotopes; $\beta^+$ decay of these isotopes emits positrons that annihilate with electrons in the detector or nearby material, yielding back‐to‐back $\gamma$‐ray pairs at $511\,\mathrm{keV}$; and high‐energy $\gamma$ rays ($E>1.022\,\mathrm{MeV}$) generated by cosmic‐ray interactions produce electron–positron pairs, whose subsequent annihilation further enhances the $511\,\mathrm{keV}$ line \citep{Diehl2018}.

\begin{figure*}[ht!]
\centering
\includegraphics[width=0.99\linewidth]{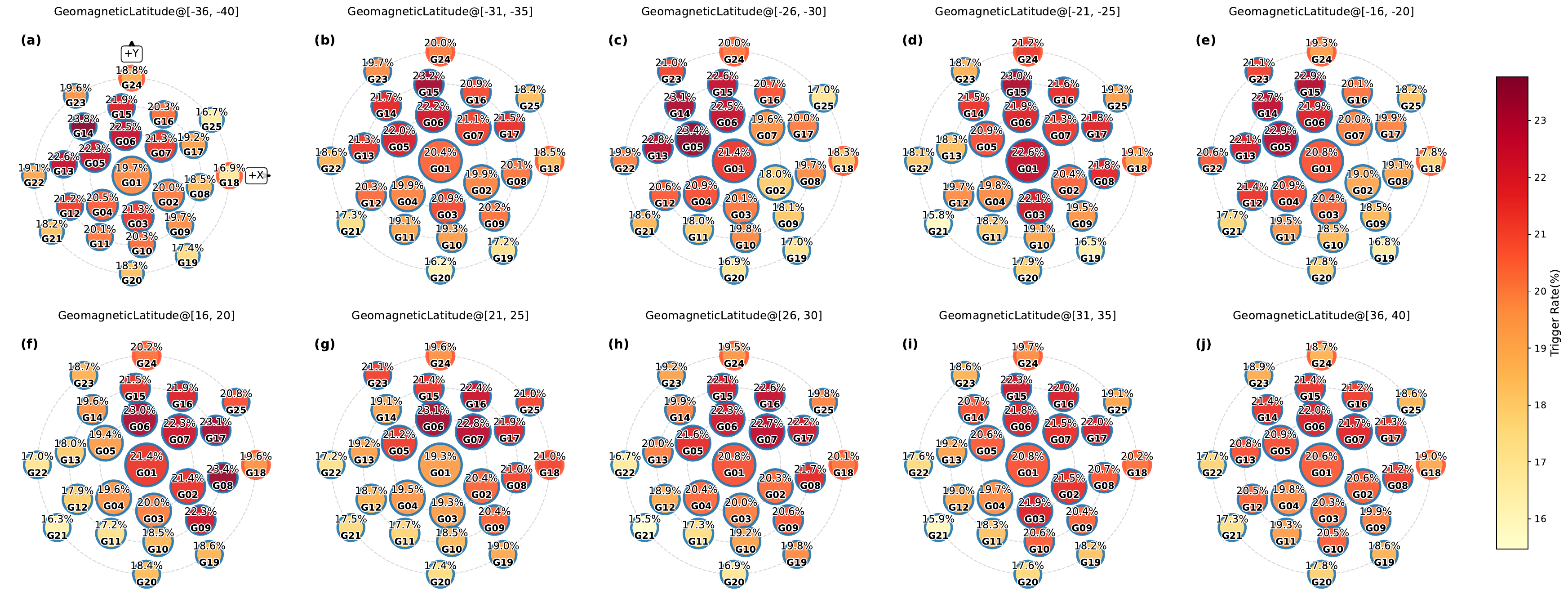}
\caption{A schematic diagram of the two-dimensional spatial distribution of the 25 GRDs (G01--G25) and their trigger rates at different geomagnetic latitudes. The numerical values (\%) displayed within each detector denote its average trigger rate, with darker shades indicating higher rates; see the color bar at right for the exact scale.}
\label{fig:fig10}
\end{figure*}

\begin{figure}[ht!]
\centering
\includegraphics[width=0.99\linewidth]{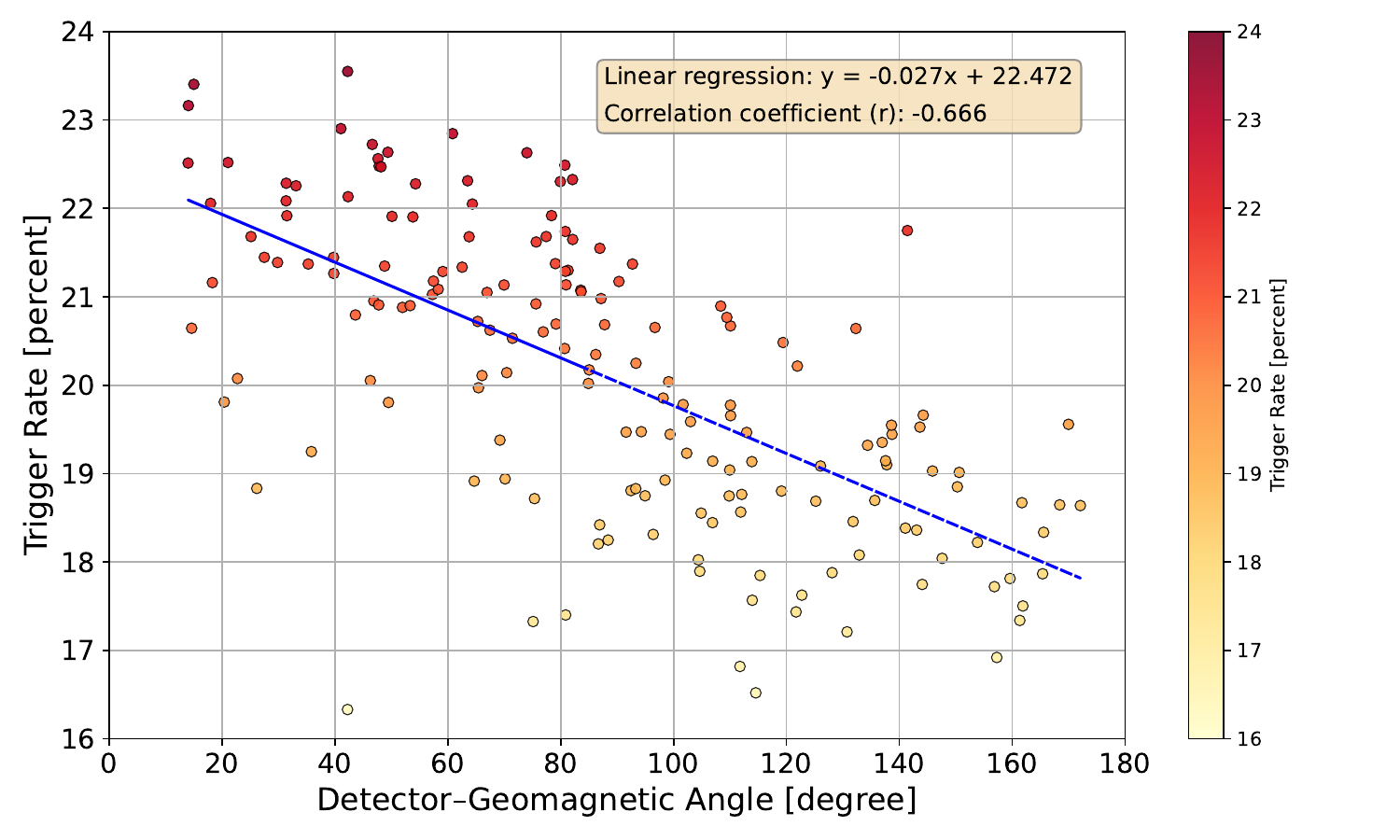}
\caption{Relationship between detector trigger rates and magnetic field line orientation. The x-axis represents the angle between GRDs and magnetic field lines, while the y-axis represents the GRD trigger rates. The blue line shows the least squares fit line, with the fitting function and correlation coefficient displayed in the upper right corner (r: indicates the strength of linear relationship between two variables).}
\label{fig:fig11}
\end{figure}

\begin{figure}[ht!]
\centering
\includegraphics[width=0.99\linewidth]{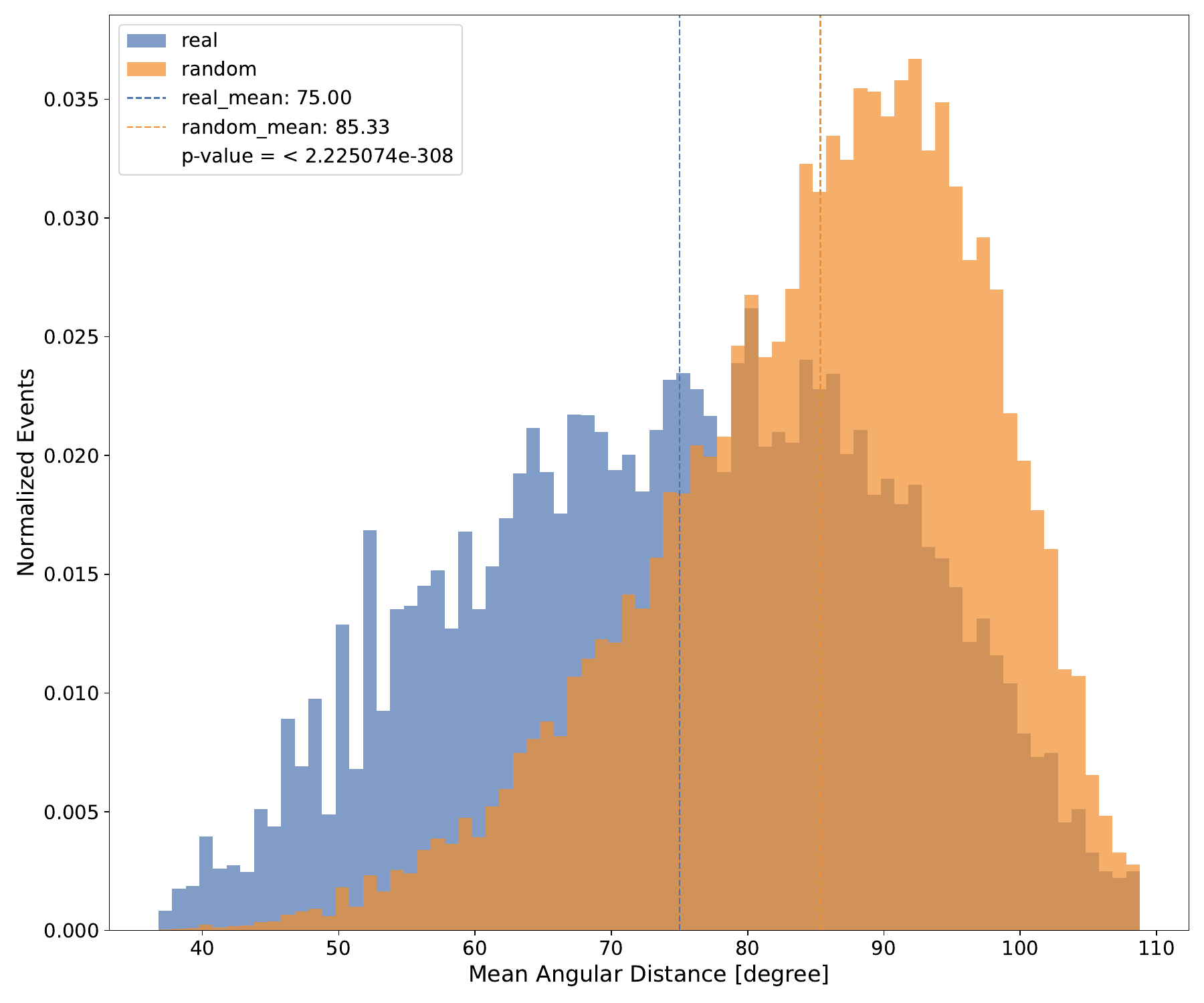}
\caption{Distribution of clustering characteristics for on-orbit data and simulated random data, where blue represents the distribution of on-orbit observational data and orange represents the distribution of randomly simulated data.}
\label{fig:fig12}
\end{figure}

\subsection{Detector-Geomagnetic Field Line Angular Correlation}

To further investigate the correlation between STE and geomagnetic field modulation, we established a quantitative analysis model relating the detector trigger rate to the angle between the detector’s normal vector and the geomagnetic field lines. This analysis computes the angle between the detector’s normal vector and the Earth’s magnetic field lines to quantitatively assess the influence of the spatial geometry between them on the detector trigger rate. Taking STE-5 as an example, we examined the trigger rate distribution across the GRD array under different geomagnetic latitude conditions. The two-dimensional spatial distribution of GRDs onboard the GECAM-B satellite and their corresponding trigger rates at various geomagnetic latitude intervals are presented in Figure \ref{fig:fig10}. The statistical analysis indicates that the outermost ring detectors (G18–G25) exhibit relatively low trigger rates across all geomagnetic latitudes, a phenomenon attributable to the spatial geometric configuration of the detector array. As edge elements, the outer ring detectors have significantly fewer spatial couplings with adjacent detectors compared to inner ring detectors, resulting in a reduced probability of receiving correlated signals. Considering this geometric effect, data from the outer ring detectors were excluded in the subsequent field-line direction dependence analysis to minimize bias due to spatial location in the analysis of the physical mechanism.

The analysis reveals a pronounced negative correlation exists between the detector trigger rate and the angle relative to geomagnetic field lines, as illustrated in Figure \ref{fig:fig11}. When the detector’s normal direction is nearly parallel to the field-line direction (small angle), the trigger rate of STE is higher; conversely, when they are nearly perpendicular or opposite (large angle), the trigger rate significantly decreases. This directional dependence has important physical significance, indicating that STE display a radial clustering along field lines, characteristic of charged particle motion, since charged particles gyrate around magnetic field lines \citep{Koskinen2022}. Therefore, the generation of STE is closely related to the trajectory of charged particles in the geomagnetic field. In contrast, neutral particles (e.g., gamma rays or neutrons) are unaffected by the magnetic field, and their propagation directions should not exhibit such a strong correlation with field-line orientation. Thus, the negative correlation between detector trigger rate and field-line angle provides direct and compelling evidence that STE have their source in charged cosmic-ray particles. This result, corroborated by the aforementioned geomagnetic latitude effect, jointly supports the association of STE with high-energy charged cosmic rays.

\subsection{Clustering of triggered Detectors}

To further analyze the physical characteristics of STE, we examined the spatial distribution of triggered detectors of each STE event. If STE were produced by independent random particle excitations, the triggered detectors would be expected to exhibit an uncorrelated, random spatial distribution. Using STE-5 as the analysis sample, we defined a quantitative clustering index by computing the pairwise spherical angular separations between each pair of the five simultaneously responding detectors in each event. This index provides a numerical descriptor of the detectors’ relative spatial configuration, effectively characterizing the degree of clustering or dispersion within each detector combination. 

We then compared the observed data with Monte Carlo simulations generated under the random-hypothesis assumption. The distribution characteristics of the observed data and Monte Carlo simulations were compared using statistical analysis, as shown in Figure \ref{fig:fig12}. The Mann–Whitney U test is used to assess whether the spatial clustering of detectors in actual events shows a systematic deviation from a random distribution \citep{Hill2018}. This nonparametric test evaluates differences in central tendency (median or location parameter) between two independent samples to evaluate whether they are drawn from populations with identical distributions, making it well suited to clustering analysis. The resulting $p$-value is extremely small, indicating that the observed data are not consistent with the random model but instead exhibit a distribution that differs significantly from it. A comparison of statistical parameters shows that the mean clustering index of the real events (\(\mathrm{real\_mean}\)) deviates markedly from the theoretical expectation for random events (\(\mathrm{random\_mean}\)), quantitatively confirming the pronounced spatial clustering of simultaneously triggered detectors in the orbital data. Specifically, in STE the detectors that record signals concurrently are predominantly spatially adjacent, whereas the simultaneous triggering of detector combinations that are spatially distant is relatively infrequent. This clustering pattern suggests that STE may have their source in particle showers: high-energy cosmic-ray particles impinging on the satellite structure interact with its material to produce cascades of secondary particles, triggering multiple neighboring detectors within a short time window. This behavior is consistent with the well-known physical phenomenon of cosmic rays generating particle showers in matter, providing direct empirical support for the hypothesis that STE arise from cosmic-ray interactions with the satellite.

\section{Non-Zenith Pointing Observation Mode}\label{section5}

Based on observations of how STE depend on magnetic field orientation, this section investigates the characteristics of their spatial source, with particular emphasis on determining whether they have their source in external signals from specific directions in the sky or from the Earth. The analysis employs a comparative approach under different satellite‐pointing conditions, examining the variation patterns of energy deposition spectra and integrated flux intensities. As mentioned above, we introduce the concept of the angle between the satellite’s Z-axis and the geocenter, denoted \(\theta\), which ranges from \(0^\circ\) (Z-axis pointing directly toward the geocenter) to \(180^\circ\) (Z-axis pointing directly away toward deep space). To eliminate systematic interference arising from variations in geomagnetic latitude, only observations collected at a fixed geomagnetic latitude are selected for this study. A sample of STE-5 is selected for analysis. Four representative angular intervals are defined: \(4^\circ\!-\!15^\circ\), \(59^\circ\!-\!81^\circ\), \(125^\circ\!-\!143^\circ\), and \(165^\circ\!-\!175^\circ\). These intervals cover the full range of satellite attitudes, from Earth-pointing to nearly zenith-pointing.

\begin{figure}[ht!]
\centering
\includegraphics[width=0.99\linewidth]{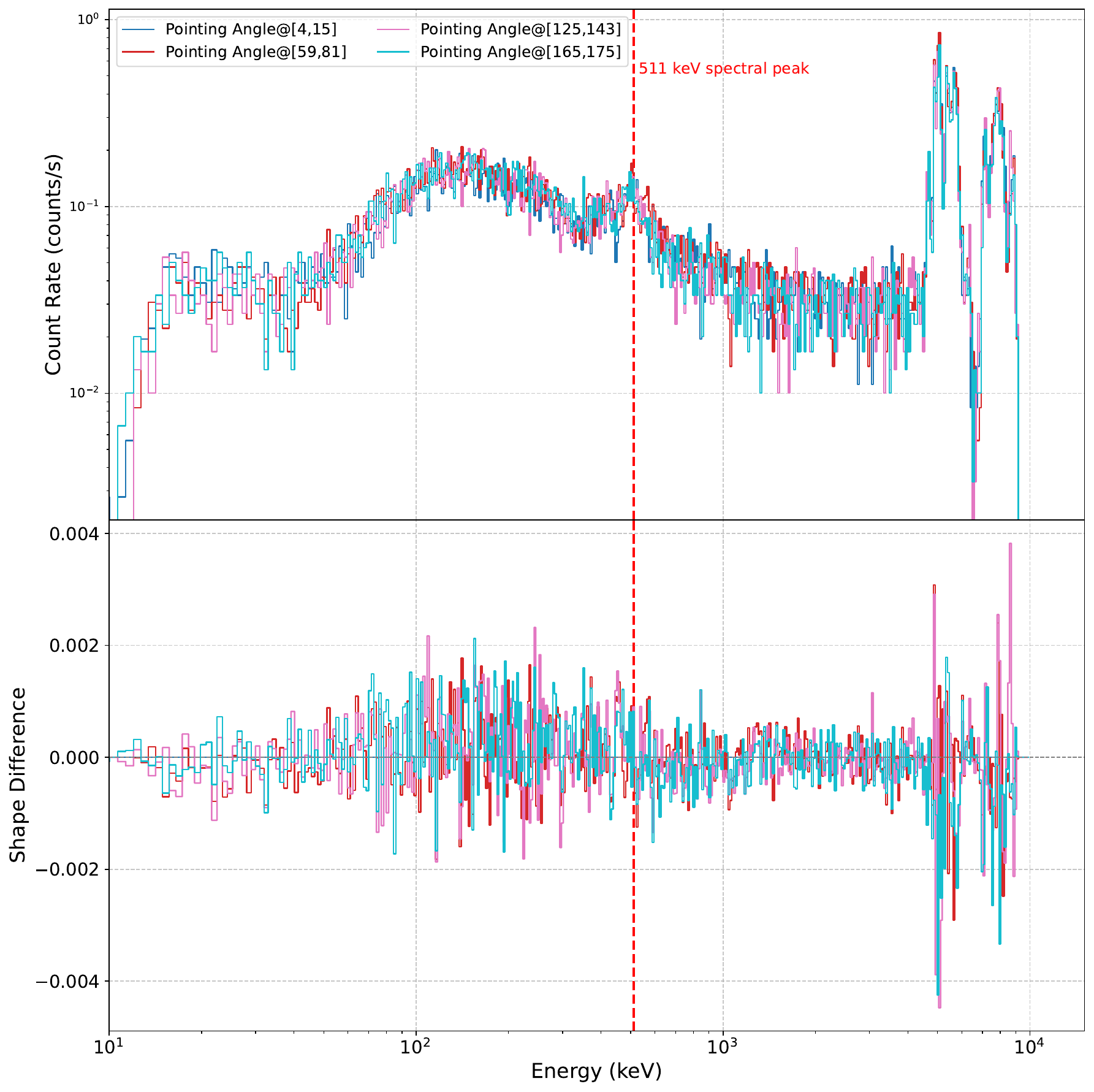}
\caption{Energy deposition spectra of STE at different pointing angles under the same geomagnetic latitude, with the red dashed line corresponding to the energy value at 511~keV.}
\label{fig:fig13}
\end{figure}

\begin{figure}[ht!]
\centering
\includegraphics[width=0.99\linewidth]{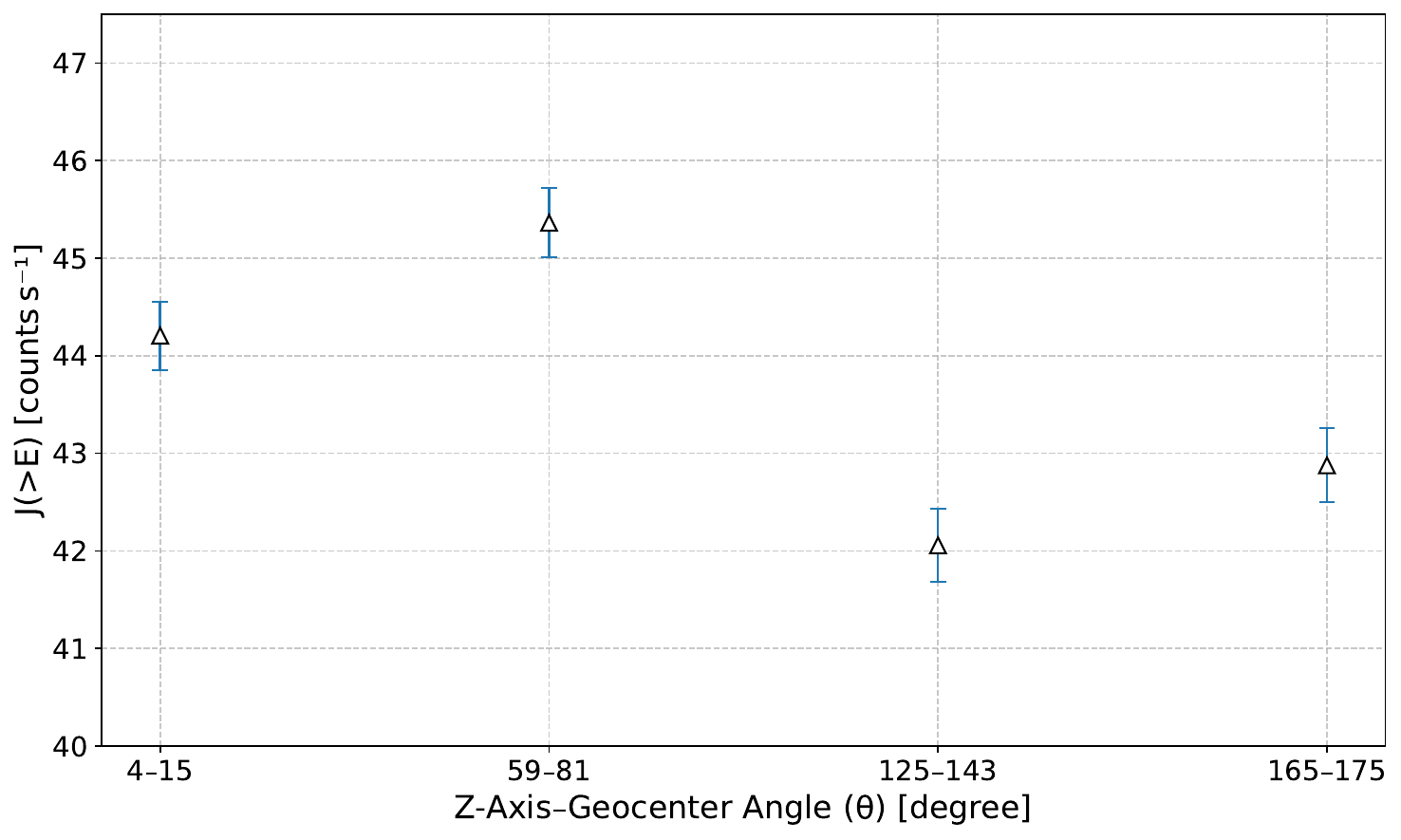}
\caption{Integral flux of STE at different pointing angles under the same geomagnetic latitude.}
\label{fig:fig14}
\end{figure}

\begin{figure}[ht!]
\centering
\includegraphics[width=0.99\linewidth]{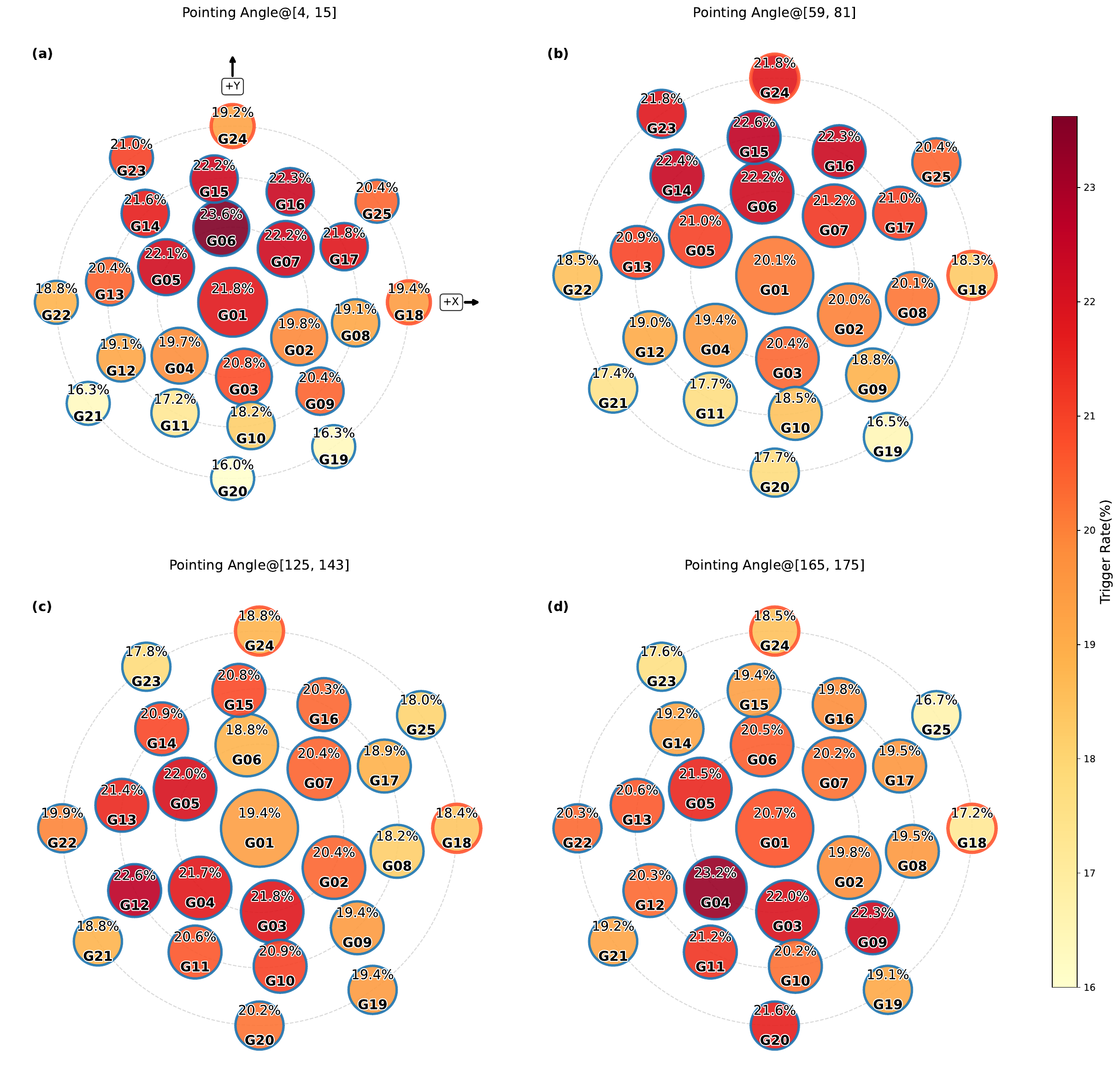}
\caption{Schematic diagram of GRD trigger rates at different pointing angles under the same geomagnetic latitude.}
\label{fig:fig15}
\end{figure}

\begin{figure}[ht!]
\centering
\includegraphics[width=0.99\linewidth]{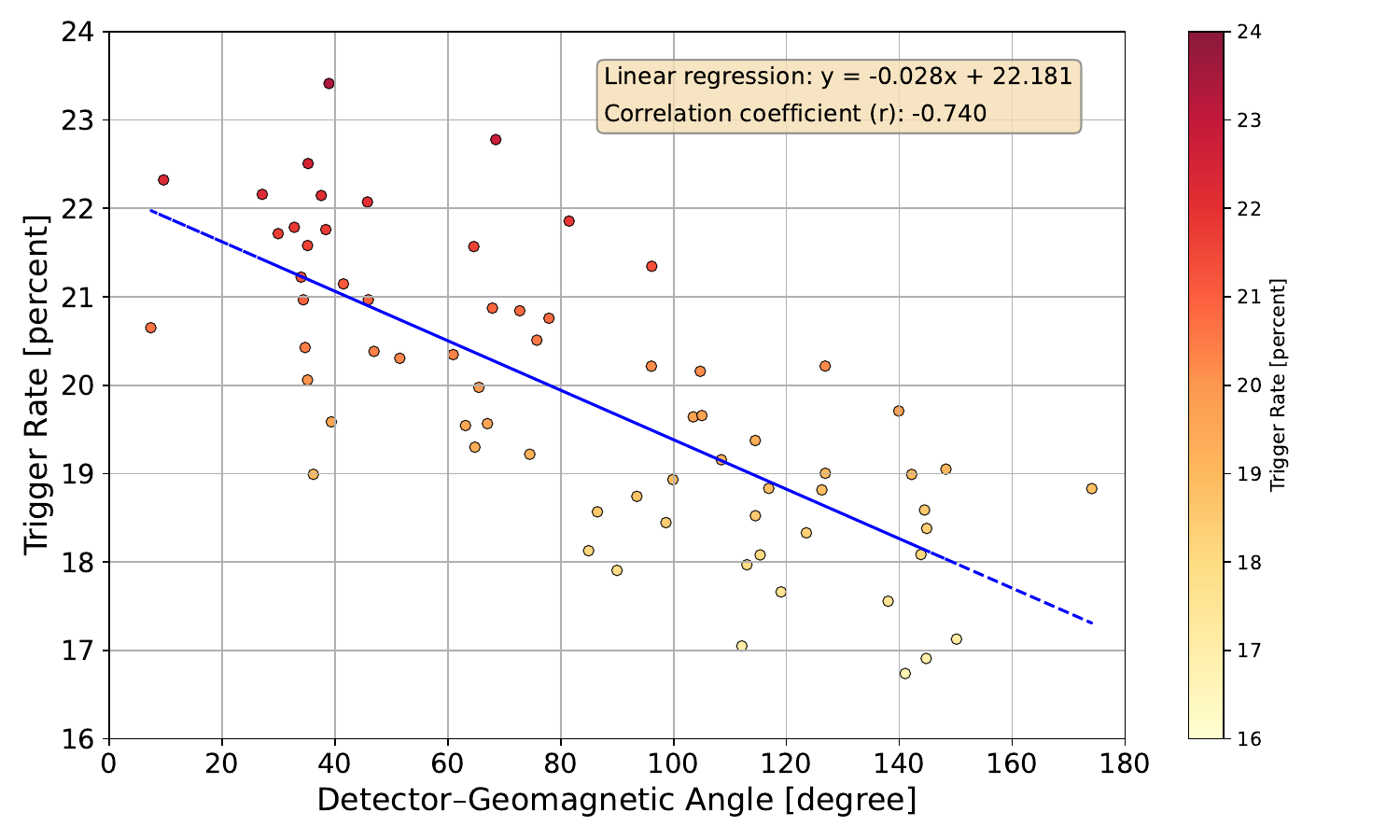}
\caption{Relationship between GRD trigger rates and magnetic field line directions at the same geomagnetic latitude.}
\label{fig:fig16}
\end{figure}

The statistical analysis reveals that the energy deposition spectra for STE maintain highly consistent shapes across all four angular intervals, with no systematic shifts observed in the relative intensity distributions across energy bands, as shown in Figure \ref{fig:fig13}. Furthermore, the integrated flux intensities of STE remain essentially constant across these angular intervals, as shown in Figure \ref{fig:fig14}. The detector trigger rates at different pointing angles (as shown in Figure \ref{fig:fig15}) and their correlation with magnetic field-line inclination angles (as shown in Figure \ref{fig:fig16}) provide additional insights into the nature of these events. The study finds a significant negative correlation between detector trigger rates and field-line inclination angles at all pointing angles. Notably, under fixed geomagnetic latitude conditions, the non-zenith pointing mode exhibits a stronger negative correlation than the zenith pointing mode. The absolute value of the correlation coefficient is markedly increased. This difference can be attributed to the modulation effect of geomagnetic latitude. By fixing the geomagnetic latitude, gradients of magnetic field strength with latitude are eliminated. Consequently, the influence of field-line inclination angle appears more purely, enhancing the statistical significance of the correlation. The energy spectrum shapes and integrated flux intensities of STE remain stable across all satellite attitudes. No systematic response patterns are observed in relation to attitude changes. This stability effectively excludes the possibility that the STE have their source in external signals fixed in a particular sky direction. If they did, changes in satellite pointing would produce corresponding variations in event rates or spectral shapes. The observed consistency suggests that these STE are more likely to exhibit isotropic distributions or arise from physical processes within the local satellite environment.

\section{Energy Spectrum Analysis}\label{section6}

\begin{figure}[ht!]
\centering
\includegraphics[width=0.99\linewidth]{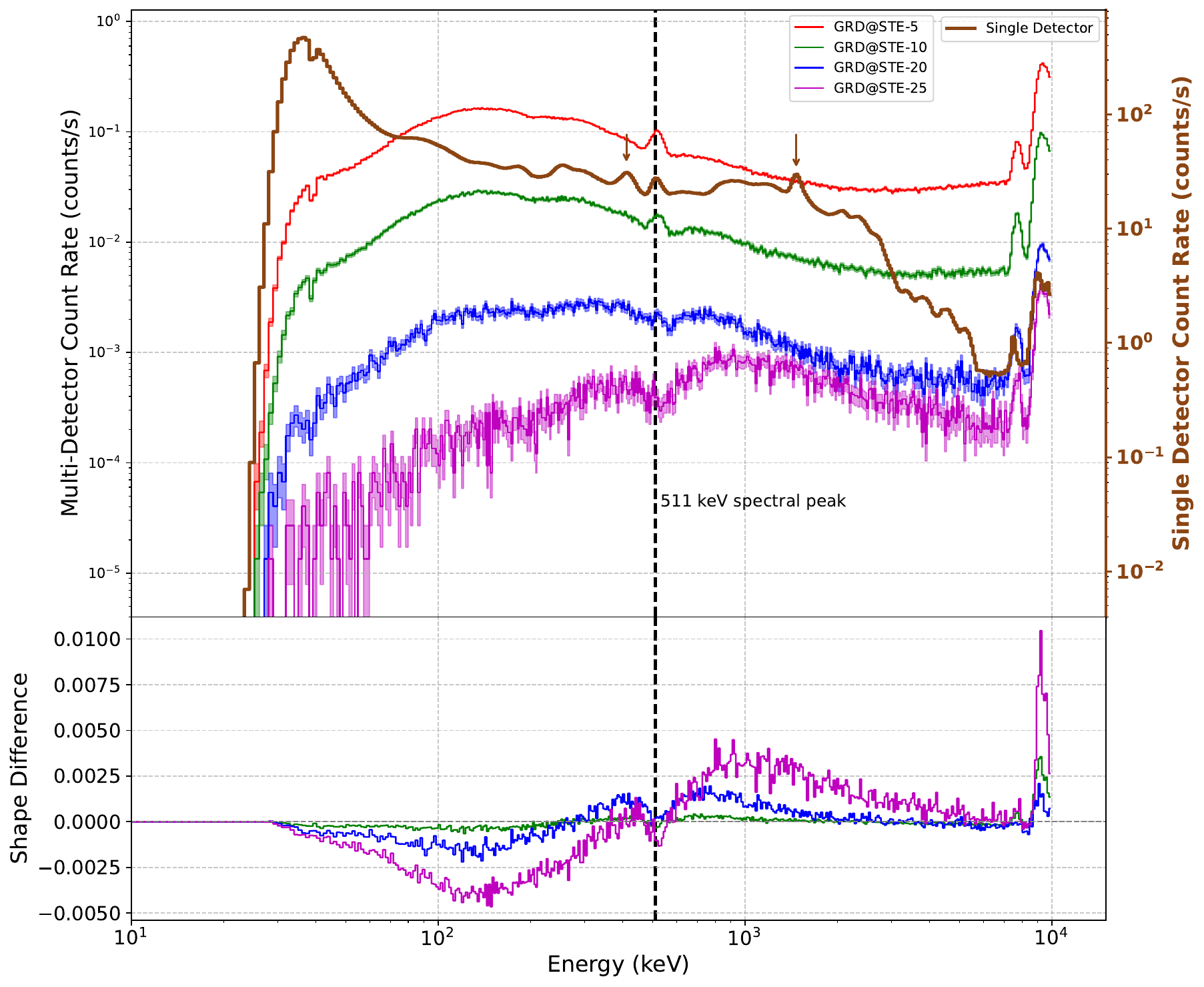}
\caption{Schematic diagram showing the evolution of single detector event spectra and STE spectra.The brown line represents the single detector event spectrum, corresponding to the right scale coordinates. From left to right, the two brown arrows indicate delayed background and intrinsic background, respectively. The red, green, blue, and magenta spectral lines correspond to STE involving 5, 10, 20, and 25 detectors, respectively. The lower panel shows the spectral shape variation characteristics of STE involving 10, 20, and 25 detectors, using the STE-5 spectrum as the reference baseline.}
\label{fig:fig17}
\end{figure}

To further elucidate the physical nature of STE, we compare the energy‐deposition spectra of single‐detector background events with those of STE for multiplicities \(N = 5,\; 10,\; 20,\; 25\). The brown curve in Figure \ref{fig:fig17} represents the spectrum of events triggering only one detector within a 0.3~\(\mu\mathrm{s}\) time window (right‐hand vertical axis). In the single‐detector spectrum, a delayed background peak appears around 410 keV, arising from the decay of radioactive nuclides activated in spacecraft materials.  Since decay is a stochastic process whose timing depends on the nuclides’ half‐lives and thus occurs with a characteristic lag, this component is termed the delayed background \citep{GuoD2020}. An intrinsic background peak at 1470~keV is attributed to \(^{138}\mathrm{La}\) decay in LaBr\(_3\) scintillators \citep{ZhangD2019,Quarati2012}, and the 511~keV feature corresponds to positron annihilation radiation induced by space radiation particles. By contrast, the spectra of STE (red, green, blue, and magenta curves in Figure \ref{fig:fig17} for \(N = 5,\; 10,\; 20,\; 25\), respectively) effectively suppress both intrinsic and delayed background contributions, confirming that the coincidence‐trigger mechanism selects non‐background physical events. The similarity of these spectra across different \(N\) values indicates a common physical source, while the increasing statistical uncertainty with higher \(N\) reflects the rarity of high‐multiplicity events. Using the \(N = 5\) spectrum as a reference, we shift its probability‐density function to a zero baseline and apply the same procedure to the other spectra in order to extract their “shape” characteristics. Analysis of the spectral‐shape evolution reveals a dual trend bifurcated at 511~keV: photon counts in the low‐energy region \( (\approx20~\mathrm{keV} \text{--} 511~\mathrm{keV}) \) decrease with increasing \(N\), whereas counts in the high‐energy region \( (511~\mathrm{keV} \text{--} \approx 10~\mathrm{MeV}) \) increase. The 511~keV annihilation peak persists across all multiplicities but diminishes in relative intensity as \(N\) grows, likely because high‐\(N\) cosmic‐ray events induce more extensive secondary particle cascades in the detector array, amplifying high‐energy counts and diluting the annihilation feature.

Overall, as \(N\) increases from 5 to 25, the spectral centroid shifts progressively toward higher energies. This evolution aligns with the energy‐deposition processes of cosmic rays in detector materials: higher multiplicities correspond to incident particles of greater energy, more complex secondary cascades, larger numbers of triggered detectors, and greater total deposited energy. The observed dual‐evolution pattern can be ascribed to energy‐threshold effects in cosmic‐ray interactions: low‐energy cosmic rays deposit energy chiefly via ionization losses in one or a few detectors, producing limited secondary particles and low‐multiplicity triggers; high‐energy cosmic rays have sufficient energy to initiate nuclear interactions, generating abundant secondary particles (neutrons, protons, mesons, etc.) that form extensive showers across the array and trigger multiple detectors simultaneously \citep{Elsevier2001}. The ubiquitous presence of the 511~keV annihilation peak underscores positron production as a hallmark of high‐energy cosmic‐ray interactions, and its dependence on \(N\) reflects the intrinsic link between cascade development and incident-particle energy.

\section{Possible origin of STE}\label{section7}

In this section we discuss the potential production mechanism of STE. We primarily evaluate two probable scenarios of high energy cosmic-rays: the atmospheric scenario and the satellite scenario. 

In the atmospheric scenario, STE are assumed to arise from secondary particles produced when high-energy cosmic rays interact with Earth’s atmosphere (primarily oxygen and nitrogen nuclei) \citep{RenY2024}. Some of these secondary particles backscatter into space and are detected by GECAM. However, our observations conflict with this hypothesis in several respects. First, if the events were sourced predominantly from the atmosphere, GECAM should record stronger signals when Earth-pointing and significantly reduced signals for zenith pointing. Instead, the energy spectra and occurrence rates of STE remain consistent across all pointing attitudes. Second, it was \citep{Alcaraz2000} demonstrated that upward-going protons from atmospheric interactions peak in flux near the magnetic equator and decrease by a factor of 2–3 as geomagnetic latitude increases from 0 to 0.3. The flux then remains relatively constant for \(0.3<\Theta_\mathrm{M}<0.8\). This behavior contradicts the observed increase in STE flux with geomagnetic latitude. Third, our observations show that STE display spatial clustering along magnetic field lines, a pattern that rules out an atmospheric neutral-particle source, since neutral particles would not be guided or focused by the geomagnetic field.

In the satellite scenario, STE have their source directly in cosmic-ray interactions in the satellite detectors or nearby materials. Incident cosmic-ray particles strike spacecraft materials and initiate secondary-particle cascades—comprising ionization products and high-energy photons—that emit quasi-isotropically and are recorded nearly simultaneously by multiple detectors. Primary cosmic rays are charged, so their trajectories are modulated by the geomagnetic rigidity cutoff. At high latitudes, a lower cutoff permits more high-energy cosmic rays to reach the satellite. This is consistent with the observed correlation between event rate and geomagnetic latitude. Cosmic rays can be trapped or deflected by the geomagnetic field, which explains the dependence of detector trigger rates for STE on the inclination of the geomagnetic field lines. The spatial clustering of simultaneously triggered detectors on the satellite platform aligns with the geometry of cascade propagation in the spacecraft material. Moreover, as detector multiplicity \(N\) increases, the energy spectrum evolves from being dominated by low-energy particles to being dominated by high-energy particles. The persistent 511~keV annihilation peak reflects the production of positrons in these cascades. Both features are in full agreement with high-energy particle cascade physics. The satellite scenario also accounts for the independence of STE characteristics from satellite pointing. Although Earth should block primary cosmic rays over an approximate \(2\pi\)~sr solid angle when Earth-pointing, the measured energy spectra and integrated fluxes remain statistically unchanged for all attitudes. This pointing-independence indicates that the particles triggering STE are not directly tied to a specific incident direction of primary cosmic rays. Instead, they are governed by secondary cascades within spacecraft materials. Even if Earth occludes certain directions, cosmic rays from unobstructed directions still produce similar cascades, maintaining the statistical stability of STE properties. Taken together, the satellite scenario self-consistently explains all observed features, whereas the atmospheric scenario exhibits multiple contradictions. We therefore conclude that STE are primarily sourced from direct interactions of cosmic rays with the satellite detectors and their surrounding materials.

\section{Discussion and Summary}\label{section8}

Taking advantage of the sub-microsecond time resolution (0.1 $\mu$s) of detectors, GECAM has the ability to record and identify the STE event, which leaves signals in multiple detectors simultaneously (actually in a time window of 0.3 $\mu$s). In this study we present the first systematic analysis of these STE detected by 25 GRDs of GECAM-B and explore the origin of these interesting events.

We find that these STE exhibit clear characteristics of a true physical event, rather than instrumental noise or random coincidence. 
The counts rate of STE increases markedly with increasing geomagnetic latitude, consistent with the cosmic-ray rigidity cutoff effect. The detector trigger rate shows a strong negative correlation with the angle between the detector's normal and the geomagnetic field line, indicating that the charged particles attributed to the production of STE propagate preferentially along those field lines. The observed spatial clustering on the satellite of triggered detectors of STE further supports this interpretation. Under different satellite pointing modes (i.e. zenith and non-zenith), the deposited energy spectrum of STE remains stable, which rules out any source in a particular sky region. As the number of simultaneously triggered detectors ($N$) increases, the spectral centroid shifts toward higher energies, and a pronounced 511 keV annihilation peak becomes apparent. Moreover, the spectral evolution of STE aligns with the cosmic-ray–induced cascade scenario. These results support the scenario that the primary source of STE is the cascade of secondary particles produced by high-energy cosmic-ray interacting with the satellite, while contribution from atmospheric secondary particles is not the dominant component. 

We note that this finding carries several important scientific implications. First, STE may be misclassified as transient high-energy phenomena such as terrestrial gamma-ray flashes\citep{Fishman1994,Dwyer2012,YiQ2025}; our analysis provides a good criteria to discriminate such false triggers. Second, STE exhibit a well-defined physical source and high synchronicity, making them very suitable for the in-flight calibration of relative timing resolution among detectors in a multi-detector system. 

More interestingly, this study preliminarily demonstrates that GECAM can detect, identify and characterize the high energy cosmic-rays through the STE. 
We have to point out that GECAM is not initially designed to detect and study cosmic-rays, and this is the by-product of the special design of GECAM instrument. For other GECAM-like all-sky monitors, cosmic-ray–induced multi-detector coincident triggers were typically regarded as background noise and filtered out. Although this approach enhances the signal-to-noise ratio for the targeted astrophysical events, it sacrifices the valuable physical information carried by the cosmic rays. 
With its unique design, GECAM enables the study of cosmic rays in the low Earth orbit. In this study, we systematically analyze these simultaneous triggers as scientifically valuable data, thereby opening a new avenue for investigating interactions between the high energy charged cosmic rays and the satellite material. 

Regarding the principle, capability and performance of cosmic-ray detection, we have to notice that GECAM differs from the dedicated cosmic-ray observatories, either the dedicated ground-based detector array which infers primary cosmic-ray information from atmospheric secondary particles, or the dedicated space-based cosmic-ray detectors which can directly measure the primary cosmic rays. 
Instead, GECAM makes use of the STE caused by the cosmic-ray cascade in the satellite materials to probe the primary cosmic-rays. 
Although such observation may not be able to yield many accurate information on the primary cosmic rays as the designated cosmic-ray facilities, it can still provide many useful information on cosmic-rays. Based on the current study in this work, GECAM is able to 
monitor the relative intensity and distribution of the charged high energy cosmic rays in the low Earth orbit. Moreover, the number of detectors triggered simultaneously, \(N\), of STE can serve as a rough indicator of the primary cosmic-ray energy. However, to achieve statistically reliable energy reconstruction, multi-factor modeling—including Monte Carlo method and spectral evolution analysis—is required.


In summary, this work not only systematically characterized the observational features of STE detected by GECAM, but also preliminarily revealed the underlying physical mechanisms of these events, 
deepening our understanding of near-Earth high-energy cosmic-ray particle environments and their interactions with satellite. As an all-sky gamma-ray monitor primarily designed to observe astrophysical transients, GECAM fulfills interesting cosmic-ray research capabilities, making GECAM as a Micro Cosmic-Ray Observatory (MICRO) in the orbit.
Future research is needed to probe cosmic-ray properties in greater depth and to optimize algorithms for distinguishing astrophysical events from cosmic-ray–induced events, thereby facilitating the study of both the high-energy transient sources and the cosmic-rays.
In addition, this work also demonstrates the feasibility of conducting cosmic-ray research with a simple design (or slight improvement in the design) of gamma-ray instrument. The design of GECAM offers references for future space-based instruments which can integrate gamma-ray and cosmic-ray observations. 

\begin{acknowledgments}
This work is supported by the National Natural Science Foundation of China 
(Grant No. 12273042, 12494572
)
, the National Key R\&D Program of China
(2022YFF0711404, 
2021YFA0718500
), 
the Strategic Priority Research Program of Chinese Academy of Sciences (Grant Nos. 
XDA15360102, XDA15360300, 
XDA30050000, 
XDB0550300
), and supported by the fund of the Key Laboratory of Cosmic Rays (Xizang University), Ministry of Education (Grant No. KLCR-2025**). 
The GECAM (Huairou-1) mission is supported by the Strategic Priority Research Program on Space Science (Grant No. XDA15360000) of the Chinese Academy of Sciences.
We thank the development and operation teams of GECAM.
\end{acknowledgments}






\bibliography{sample701}{}
\bibliographystyle{aasjournalv7}



\end{document}